\documentclass[aip,jcp,preprint]{revtex4-1}
\usepackage{amsmath,amsfonts,amssymb,amsthm}
\usepackage{placeins}
\usepackage{tabularx,multirow}
\usepackage[T1]{fontenc}
\usepackage{graphicx}
\usepackage[english]{babel}
\usepackage{xcolor}
\usepackage{natbib}
\usepackage{amsmath}
\usepackage{amsfonts}
\usepackage{algcompatible}
\usepackage{newfloat}

\algblockdefx[PROCEDURE]{PROCEDURE}{ENDPROCEDURE}[2][]{#1\textbf{procedure} #2}{\textbf{end procedure}}
\DeclareFloatingEnvironment[
    fileext=loa,
    listname=List of Algorithms,
    name=ALGORITHM,
    placement=tbhp,
]{algorithm}
\usepackage{amssymb}%
\usepackage{wasysym}
\usepackage{bm}
\newcommand\pseudodot[1]{}
\providecommand{\U}[1]{\protect\rule{.1in}{.1in}}
\graphicspath{{./Figures/}{D:/Papers/Konstantin/EIE_extended_space/Figures/}
{C:/Users/Jiri/Dropbox/Papers/Chemistry_papers/2016/EIE_stochastic_DE/Figures/}{./}
}
\makeatletter
\providecommand{\tabularnewline}{\\}

\@ifundefined{textcolor}{}
{
\definecolor{BLACK}{gray}{0}
\definecolor{WHITE}{gray}{1}
\definecolor{RED}{rgb}{1,0,0}
\definecolor{GREEN}{rgb}{0,1,0}
\definecolor{BLUE}{rgb}{0,0,1}
\definecolor{CYAN}{cmyk}{1,0,0,0}
\definecolor{MAGENTA}{cmyk}{0,1,0,0}
\definecolor{YELLOW}{cmyk}{0,0,1,0}
}
\makeatother
\newcommand{\rot}[1]{\begin{tabular}{c}\rotatebox{90}{\begin{tabular}{c}#1\end{tabular}}\end{tabular}}
\newcommand{\pz}{\phantom{0}}
\begin{document}
\title{A combined on-the-fly/interpolation procedure for evaluating energy values needed in molecular simulations}
\author{Konstantin Karandashev}
\email{konstantin.karandashev@alumni.epfl.ch}
\author{Ji\v{r}\'{\i} Van\'{\i}\v{c}ek}
\email{jiri.vanicek@epfl.ch}
\affiliation{Laboratory of Theoretical Physical Chemistry, Institut des Sciences et
Ing\'{e}nierie Chimiques, Ecole Polytechnique F\'{e}d\'{e}rale de Lausanne
(EPFL), CH-1015, Lausanne, Switzerland}
\date{\today}

\begin{abstract}

We propose an algorithm for molecular dynamics or Monte Carlo simulations that uses an interpolation procedure to estimate potential energy values from energies and gradients evaluated previously at points of a simplicial mesh. We chose an interpolation procedure which is exact for harmonic systems and considered two possible mesh types: Delaunay triangulation and an alternative anisotropic triangulation designed to improve performance in anharmonic systems. The mesh is generated and updated on the fly during the simulation.  The procedure is tested on two-dimensional quartic oscillators and on the path integral Monte Carlo evaluation of $ \mathrm{HCN}/\mathrm{DCN}$ equilibrium isotope effect.

\end{abstract}
\maketitle

\section{Introduction}

Accurate evaluation of the Born-Oppenheimer potential energy surface of a molecular system is essential for predicting its dynamical and equilibrium properties. Numerous advances in the algorithms used for the problem\cite{Song_Martinez:2016,Jeanmairet_Alavi:2017} combined with increasing computational power available to researchers have made it possible to combine on-the-fly {\it ab initio} evaluation of the potential energy even with path integral\cite{Perez_Lilienfeld:2010,Gasparotto_Ceriotti:2016,PerezDeTudela_Marx:2017} or semiclassical\cite{Worth_Burghardt:2004,Ianconescu_Pollak:2013,Curchod_Martinez:2018,Spinlove_Worth:2018,Patoz_Vanicek:2018,Micciarelli_Ceotto:2019} dynamics algorithms. Unfortunately, such approaches are still computationally expensive and, for long simulations requiring a very large number of potential energy values in the same region of configuration space, it is reasonable to instead generate a mesh of points at which accurate {\it ab initio} calculations are made and then fit a function to reproduce their potential energy values or some other potential quantities that become bottlenecks of the calculation, such as Hessians of the potential energy.\cite{Laude_Richardson:2018,Conte_Ceotto:2019} For that purpose, a multitude of methods has been proposed, from modified Shepard interpolation\cite{Ischtwan_Collins:1994,Frankcombe_Worth:2010,Evenhuis_Martinez:2011,Kim_Rhee:2016} to more sophisticated approaches,\cite{Huang_Bowman:2005} including those based on interpolating moving least squares,\cite{Guo_Minkoff:2004,Dawes_Carrington:2010} Gaussian process regression,\cite{Alborzpour_Habershon:2016,Richings_Habershon:2017} and neural networks.\cite{Blank_Doren:1995,Manzhos_Carrington:2006,Malshe_Komanduri:2010,Gastegger_Marquetand:2017}

We aimed for a procedure that would interpolate energies from stored data evaluated at points of a simplicial mesh and that would be comparable to Shepard interpolation in terms of simplicity and generality. To that end, we investigated interpolation from points of the mesh that constitute a simplex containing the point of interest, an approach already applied to some lower-dimensional systems.\cite{Salazar_Bell:1998,Salazar:2002} Compared to Shepard interpolation, the downside of this approach is the necessity to generate a triangulation for the mesh, whose size grows very fast with the number of dimensions,\cite{Hornus_Boissonnat:2008} but the upside is the logarithmic scaling of the interpolation procedure with the number of mesh points as well as an extra order of accuracy for a given number of derivatives available at the mesh points. In comparison to the method of Ref.~\onlinecite{Salazar:2002}, the main differences in the approach presented here are an alternative triangulation of the mesh and a different choice of the interpolant, along with a procedure for updating the mesh during the simulation.

The theory behind the proposed algorithm is explained in Sec.~\ref{sec:Theory}, while Sec.~\ref{sec:Applications} 
presents numerical tests for model anharmonic potentials and for the $ \mathrm{HCN/DCN}$ equilibrium isotope effect. While we focus on classical Monte Carlo and path integral Monte Carlo applications, similar interpolation procedures can be also used with molecular dynamics or path integral molecular dynamics methods.

\section{Theory}
\label{sec:Theory}

Running a Monte Carlo simulation requires knowledge of the potential energy function $ V(\mathbf{r})$, where $ \mathbf{r}$ is the $ D$-dimensional vector of system's internal coordinates. Let us assume that we can access a number of previously stored points together with their potential energy and gradient values as well as a triangulation of their mesh; we want to use that information to estimate the potential energy value at a new point $\tilde{\mathbf{r}}$. If the mesh currently contains fewer than $ D+1$ points, $ V(\tilde{\mathbf{r}})$ is evaluated exactly and $ \tilde{\mathbf{r}}$ is added to the mesh, which will be triangulated (in the only possible way) once $ D+1$ points have been added. If the mesh has already been triangulated the following algorithm is used for estimating $ V(\tilde{\mathbf{r}})$:
\begin{enumerate}
 \item Find the simplex $ \tilde{\mathcal{S}}$ containing $ \tilde{\mathbf{r}}$ or verify that $ \tilde{\mathbf{r}}$ lies outside the convex hull $ \mathcal{C}_{\mathrm{mesh}}$ of all mesh points.
 \item If $ \tilde{\mathcal{S}}$ was found, calculate the value of the interpolant $ \tilde{V}(\tilde{\mathbf{r}})$ and estimate whether the interpolation error $ |\tilde{V}(\tilde{\mathbf{r}})-V(\tilde{\mathbf{r}})|$ is below a predefined threshold.
 \item If $ \tilde{\mathbf{r}}\notin\mathcal{C}_{\mathrm{mesh}}$ or if $ V(\tilde{\mathbf{r}})$ cannot be estimated with sufficient accuracy, add more points to the mesh to allow for an accurate estimate of $ V(\tilde{\mathbf{r}})$.
\end{enumerate}
We will discuss each part of the algorithm separately in the following subsections.

\subsection{Interpolation procedure and reliability estimate}
\label{subsec:interpolant}

Suppose $ \tilde{\mathbf{r}}$ is inside simplex $ \tilde{\mathcal{S}}$ with vertices $ \mathbf{r}_{\tilde{\mathcal{S}}}^{j}$ ($ j=1,\ldots,D+1$) and we want to estimate $ V(\tilde{\mathbf{r}})$ based on the values of the energy and its gradient at the $ D+1$ points $ \mathbf{r}_{\tilde{\mathcal{S}}}^{j}$. Previously, Clough-Tocher interpolants\cite{Clough_Tocher:1966,Alfeld:1984} were used for the problem in up to three dimensions;\cite{Salazar_Bell:1998,Salazar:2002} these interpolation schemes are exact for cubic potentials and have derivatives that are continuous up to the second order, but they have two disadvantages: they use Hessians, whose evaluation increases enormously the cost of an {\it ab initio} calculation, and their generalization to higher-dimensional systems is not straightforward. Perpendicular interpolation\cite{Alfeld:1985} is another powerful approach which, for an arbitrary number of dimensions and an arbitrary number of derivatives $ q$ available for all vertices, produces an interpolant that is exact for a polynomial of order $q+1$ and that has $ q$ continuous derivatives; however it scales exponentially with dimensionality $ D$, making potential applications to higher-dimensional systems problematic. In this work we used an interpolant that exhibits a better scaling with $ D$ at the cost of having discontinuous derivatives. (If this is a problem, interpolants of Ref.~\onlinecite{Alfeld:1985} should be used instead.) To define this interpolant we introduce barycentric coordinates $ \lambda_{j}$ ($ j=1,\ldots,D+1$) of $ \tilde{\mathbf{r}}$, which are defined by the $ D+1$ equations
\begin{equation}
\begin{split}
 \sum_{j=1}^{D+1}\lambda_{j}\mathbf{r}^{j}_{\tilde{\mathcal{S}}}=\tilde{\mathbf{r}},\\
 \sum_{j=1}^{D+1}\lambda_{j}=1.
\end{split}
 \label{eq:barycent_coord_def}
\end{equation}
The interpolant we propose is defined in terms of ``partial'' interpolants $ \tilde{V}^{j}$
\begin{equation}
 \tilde{V}^{j}(\tilde{\mathbf{r}})=V(\mathbf{r}_{\tilde{\mathcal{S}}}^{j})+
 \frac{1}{2}[\nabla V(\mathbf{r}_{\tilde{\mathcal{S}}}^{j})+\sum_{j^{\prime}=1}^{D+1}\lambda_{j^{\prime}}\nabla V(\mathbf{r}_{\tilde{\mathcal{S}}}^{j^{\prime}})]\cdot(\tilde{\mathbf{r}}-\mathbf{r}_{\tilde{\mathcal{S}}}^{j}),
 \label{eq:tilde_V_j}
\end{equation}
all of which are exact for quadratic potentials. One way to combine them into a single interpolant symmetric with respect to vertex permutations is
\begin{align}
\begin{split}
 \sum_{j=1}^{D+1}\lambda_{j}\tilde{V}^{j}(\tilde{\mathbf{r}})=&\sum_{j=1}^{D+1}\lambda_{j}\left[V(\mathbf{r}_{\tilde{S}}^{j})+\frac{1}{2}\nabla V(\mathbf{r}_{\tilde{S}}^{j})\cdot(\tilde{\mathbf{r}}-\mathbf{r}_{\tilde{\mathcal{S}}}^{j})\right]\\
 &+\frac{1}{2}\left[\sum_{j^{\prime}=1}^{D+1}\lambda_{j^{\prime}}\nabla V(\mathbf{r}_{\tilde{\mathcal{S}}}^{j^{\prime}})\right]\cdot\left[\sum_{j=1}^{D+1}\lambda_{j}(\tilde{\mathbf{r}}-\mathbf{r}_{\tilde{\mathcal{S}}}^{j})\right],
 \end{split}\\
 =&\sum_{j=1}^{D+1}\lambda_{j}\left[V(\mathbf{r}_{\tilde{S}}^{j})+\frac{1}{2}\nabla V(\mathbf{r}_{\tilde{S}}^{j})\cdot(\tilde{\mathbf{r}}-\mathbf{r}_{\tilde{\mathcal{S}}}^{j})\right],
\end{align}
which is an interpolant proposed in Ref.~\onlinecite{Dell'Accio_Zerroudi:2018} (based on Refs.~\onlinecite{Xuli:2003} and~\onlinecite{Guessab_Schmeisser:2006}). The term on the second line is zero because the second factor is zero. This combination of $ \tilde{V}^{j}$, however, would not reproduce potential energy gradient at the vertices, which is a waste since each $ \tilde{V}^{j}$ reproduces the gradient at vertex $ j$. An alternative expression that does reproduce gradients at all vertices is
\begin{equation}
 \tilde{V}(\tilde{\mathbf{r}})=\frac{\sum_{j=1}^{D+1}\lambda_{j}^{2}\tilde{V}^{j}(\tilde{\mathbf{r}})}{\sum_{j=1}^{D+1}\lambda_{j}^{2}}.
  \label{eq:tilde_V}
\end{equation}

It is impossible to get a reliable estimate of the interpolation error without any knowledge of the third derivatives of $ V(\mathbf{r})$ in the simplex and application of Bayesian approaches as in Shepard interpolation\cite{Bettens_Collins:1999} is complicated by $ \tilde{V}^{j}(\tilde{\mathbf{r}})$ containing data from all vertices of the simplex at once. One exception is the one-dimensional case, where defining
\begin{equation}
 \delta V(\tilde{\mathbf{r}})=\max_{j=1,\ldots,D+1}|\tilde{V}(\tilde{\mathbf{r}})-\tilde{V}^{j}(\tilde{\mathbf{r}})|
 \label{eq:delta_V_definition}
 \end{equation}
(with $ D=1$)  yields an exact estimate $ |\tilde{V}(\tilde{\mathbf{r}})-V(\tilde{\mathbf{r}})|\leq \delta V(\tilde{\mathbf{r}})$. The estimate seems to perform qualitatively correctly for a large number of higher-dimensional potentials as well, so we decided to deem the interpolation result reliable if $\delta V(\tilde{\mathbf{r}})$ were below some predetermined threshold $\delta V_{\mathrm{max}} $. This makes $\delta V_{\mathrm{max}} $ a parameter whose only relation to interpolation error is that both are proportional to the magnitude of third derivatives in a small enough simplex. The lack of a more precise relation between the two quantities forced us to estimate the {\it exact} interpolation error for a small number of randomly chosen interpolation results, determining whether the chosen $\delta V_{\mathrm{max}} $ is adequate. For small enough simplices the leading contributions to both the interpolation error and $ \delta V(\tilde{\mathbf{r}})$ depend linearly on the tensor of third derivatives of potential energy, so one possible rule of thumb for fixing an unacceptable interpolation error would be a proportional decrease of $\delta V_{\mathrm{max}} $, e.g. halving $\delta V_{\mathrm{max}} $ if root mean square error (RMSE) of interpolation should be halved.

\subsection{Updating the mesh and its triangulation}
\label{subsec:mesh_update}

If we either find $ \tilde{\mathbf{r}}$ to be outside the convex hull $ \mathcal{C}_{\mathrm{mesh}}$ or that $ \delta V\geq\delta V_{\mathrm{max}} $, we add a carefully chosen point $ \mathbf{r}_{\mathrm{add}}$ to the mesh and update the triangulation. Before describing the algorithm, let us introduce several definitions. Firstly, the boundary of $\mathcal{C}_{\mathrm{mesh}}$ is a set of faces referred to as $ \bm{\mathcal{F}}_{\mathrm{mesh}}$. Secondly, we will often use the signed distance $ n_{\mathcal{F}}(\mathbf{r})$ from the plane containing the face $\mathcal{F}\in\bm{\mathcal{F}}_{\mathrm{mesh}}$, with sign defined to be nonnegative at the mesh points. Lastly, both systems that will be considered in Sec.~\ref{sec:Applications} have certain restrictions on the values $ \mathbf{r}$ can take: in the symmetric two-dimensional quartic oscillator, the symmetry makes it possible to consider values of coordinates in only one quadrant (e.g., both $ x$ and $ y$ non-negative), while in HCN our choice of internal coordinates implies that two of them only take non-negative values. We thus consider a situation where $ \mathbf{r}$ needs to satisfy one or more linear constraints of the form
\begin{equation}
 \mathbf{v}_{i}\cdot\mathbf{r}-c_{i}\geq 0,
 \label{eq:constranints}
\end{equation}
where $ i $ is an index of the constraint, $ c_{i}$ is a scalar constant, and $ \mathbf{v}_{i}$ is a vector constant.

When $ \tilde{\mathbf{r}}$ was inside $ \mathcal{C}_{\mathrm{mesh}}$, we found that adding a point $ \mathbf{r}_{\mathrm{add}}=\tilde{\mathbf{r}}$ to the mesh worked well enough. However, when $ \tilde{\mathbf{r}}$ was outside $ \mathcal{C}_{\mathrm{mesh}}$, we ``pushed'' $ \mathbf{r}_{\mathrm{add}}$ further out, i.e., chose $ \mathbf{r}_{\mathrm{add}}$ further away from $ \mathcal{C}_{\mathrm{mesh}}$ than $ \tilde{\mathbf{r}}$, primarily to avoid creating nearly singular simplices. The procedure, referred to as ``outward push,'' works as follows:
\begin{enumerate}
 \item Find $ \mathcal{F}_{\mathrm{min}}\in\bm{\mathcal{F}}_{\mathrm{mesh}}$ that minimizes $ n_{\mathcal{F}}(\tilde{\mathbf{r}})$.
 \item Set $ \mathbf{r}_{\mathrm{add}}=\tilde{\mathbf{r}}-c_{\mathrm{push}} \nabla n_{\mathcal{F}_{\mathrm{min}}}$, where $c_{\mathrm{push}}$ is some constant.
 \item If, for some $ i $, $ \mathbf{v}_{i}\cdot\mathbf{r}_{\mathrm{add}}-c_{i}<0$, then replace $ \mathbf{r}_{\mathrm{add}}$ with
 \begin{equation}
\mathbf{r}_{\mathrm{add}}-\frac{\mathbf{v}_{i}}{|\mathbf{v}_{i}|^{2}}(\mathbf{v}_{i}\cdot\mathbf{r}_{\mathrm{add}}-c_{i}).
 \end{equation}
 \end{enumerate}
Step~3 is designed to move the mesh point (pushed away in Steps~1-2) onto one of the ``constraining surfaces'' defined by
\begin{equation}
 \mathbf{v}_{i}\cdot\mathbf{r}-c_{i}=0,
\label{eq:constraining_surface}
\end{equation}
instead of rejecting a move that would violate the constraint. Several such rejections would lead to a mesh that would approach
infinitely close to one of the constraining surfaces during the simulation, as illustrated in Subsec.~\ref{subsec:sqo_applications}.

 Once $ \mathbf{r}_{\mathrm{add}}$ is chosen, we need to update the triangulation of the mesh. Here we only present the main ideas of the employed algorithms; the details are in the Appendix. The starting point of this work was using Delaunay triangulation\cite{Lawson:1986} following its previous successful applications.\cite{Salazar_Bell:1998,Salazar:2002} There exist several algorithms\cite{Hornus_Boissonnat:2008} that update a Delaunay triangulation at a cost that does not increase with the number of simplices. The approach outlined in this subsection uses Lawson flips\cite{Lawson:1986} defined using ``parabolic lifting''\cite{Edelsbrunner:2000} instead of the more conventional empty circumsphere test.\cite{Lawson:1986,Edelsbrunner:2000} One considers sets of $ D+2$ points, which can be triangulated at most in two different ways.\cite{Lawson:1986} Whenever such a set is already triangulated using simplex array $ \bm{\mathcal{S}}$ and an alternative triangulation $ \bm{\mathcal{S}}^{\prime}$ is available, one compares the values $ G(\bm{\mathcal{S}})$ and $ G(\bm{\mathcal{S}}^{\prime})$, where the function $ G$ is defined as
 \begin{equation}
   G(\bm{\mathcal{S}})=\sum_{\mathcal{S}\in\bm{\mathcal{S}}}g(\mathcal{S})v(\mathcal{S})\label{eq:flip_functional}
  \end{equation}
 and where $ v(\mathcal{S})$ is the volume of simplex $ \mathcal{S}$. The choice of the cost function $ g(\mathcal{S})$ that yields Delaunay triangulation is
\begin{equation}
 g_{\mathrm{Delaunay}}(\mathcal{S})=\sum_{j=1}^{D+1}|\mathbf{r}^{j}_{\mathcal{S}}|^{2}.
\end{equation}
If $ G_{\mathrm{Delaunay}}(\bm{\mathcal{S}})>G_{\mathrm{Delaunay}}(\bm{\mathcal{S}}^{\prime})$, which is equivalent to $ \bm{\mathcal{S}}$ failing the empty circumsphere test used to define Delaunay triangulation,\cite{Lawson:1986,Edelsbrunner:2000} the simplices of $ \bm{\mathcal{S}}$ are replaced with those of $ \bm{\mathcal{S}}^{\prime}$.

One performs Lawson flips until they fail to change the triangulation regardless of the initial $ \bm{\mathcal{S}}$. Since each Lawson flip decreases $G_{\mathrm{Delaunay}}(\bm{\mathcal{S}}_{\mathrm{mesh}})$, where $\bm{\mathcal{S}}_{\mathrm{mesh}}$ is the array of all simplices in $ \mathcal{C}_{\mathrm{mesh}}$, the algorithm is bound to stop at a certain point, and it can be proven\cite{Lawson:1986} that the resulting final triangulation is unique to the mesh. It can also be shown that $ G(\bm{\mathcal{S}})\neq G(\bm{\mathcal{S}}^{\prime})$ for the two triangulations of $ D+2 $ points unless the points lie on a sphere or in a hyperplane; treatment of these singular cases is discussed in the Appendix.

The expression for $ g_{\mathrm{Delaunay}}(\mathcal{S})$ underlines one problem with Delaunay triangulation: it treats all dimensions equivalently, necessitating a choice of internal coordinates that makes properties of $ V(\mathbf{r})$ approximately isotropic, which tends to be non-trivial. A Bayesian approach to bypassing the problem for Shepard interpolation is discussed in Ref.~\onlinecite{Bettens_Collins:1999}, while for simplex interpolation one can use higher-order derivatives to define a Riemannian metric\cite{Mirebeau:2010} that can then be used to construct the triangulation optimal for the current interpolation procedure.\cite{Bossen_Heckbert:1996,Shimada_Itoh:2000,Boissonnat_Wintraecken:2017} Unfortunately, for quadratic interpolation the latter option would involve calculating third derivatives of the potential, which is rather expensive; therefore, we instead used Lawson flips with a modified $ g(\mathcal{S})$. Obviously, the procedure still stops at a certain triangulation regardless of the choice of $ g(\mathcal{S})$, even though we will not be able to guarantee the triangulation's uniqueness without restrictions on the potential $ V(\mathbf{r})$. The anisotropic $ g(\mathcal{S})$ proposed in this work was
\begin{equation}
g_{\mathrm{anisotr}}(\mathcal{S})=\max_{j^{\prime},j^{\prime\prime}=1,\ldots, D+1}\left|V(\mathbf{r}^{j^{\prime}}_{\mathcal{S}})-V(\mathbf{r}^{j^{\prime\prime}}_{\mathcal{S}})-\frac{[\nabla V(\mathbf{r}^{j^{\prime}}_{\mathcal{S}})+ \nabla V(\mathbf{r}^{j^{\prime\prime}}_{\mathcal{S}})]\cdot(\mathbf{r}^{j^{\prime}}_{\mathcal{S}}-\mathbf{r}^{j^{\prime\prime}}_{\mathcal{S}})}{2}\right|
\label{eq:f_anharm} 
\end{equation}
which is a qualitative estimate of the upper bound for interpolation error in a given simplex; unlike $ g_{\mathrm{Delaunay}}$, $g_{\mathrm{anisotr}}$ is invariant with respect to linear transformations of coordinates. To avoid entering infinite loops for cases when $ G(\bm{\mathcal{S}})=G(\bm{\mathcal{S}}^{\prime})$, we modify the flipping criterion to be $ G(\bm{\mathcal{S}})-G(\bm{\mathcal{S}}^{\prime})>\delta G_{\mathrm{min}}$, where $ \delta G_{\mathrm{min}}$ is a small predefined parameter.

$ g_{\mathrm{anisotr}}$ should be applicable for any simplex interpolant which uses potential and its gradient and is exact for quadratic potentials. To understand why quadratic interpolation is special in this context, consider interpolating from $ D+1$ vertices with $ q$ derivatives available, which in general can yield an interpolant exact for polynomials up to degree $q+1$. In the special case of $ D=1$, $ q$ polynomials of degree $ q+1$ can be constructed from $ 2(q+1)$ parameters available; for $ q=1$ $ g_{\mathrm{anisotr}}$ arises naturally as the degree to which the two cubic polynomials disagree. The $ q=1$ case is special because for larger values of $ q$ and more than two polynomials of degree $(q+1)$, several analogues of $ g_{\mathrm{anisotr}}$ are possible, while for $q=0$ the (linear) interpolant is uniquely defined, making it impossible to define a similar $ g_{\mathrm{anisotr}}$.

From now on, the triangulation that results from using $ g_{\mathrm{anisotr}}$ with Lawson flips will be referred to as ``anisotropic triangulation''.

\subsection{Search for the simplex}
\label{subsec:simplex_search}

The last task is finding the simplex $\tilde{\mathcal{S}}$ that contains $\tilde{\mathbf{r}}$. We used {\it stochastic walk}\cite{Devillers_Teillaud:2002} that iteratively updates $\tilde{\mathcal{S}}$ from an initial guess $ \tilde{\mathcal{S}}_{\mathrm{init}}$ by calculating barycentric coordinates $\lambda_{j} $~[Eq.~(\ref{eq:barycent_coord_def})], then terminating the search if all $ \lambda_{j}$ are positive or if a negative $ \lambda_{j}$ corresponds to a face which is also a face of $ \mathcal{C}_{\mathrm{mesh}}$, and otherwise obtaining the next $ \mathcal{S}$ as the simplex across a randomly chosen face corresponding to a negative $ \lambda_{j}$. The procedure is guaranteed to find $\tilde{\mathcal{S}}$ regardless of the triangulation used.\cite{Devillers_Teillaud:2002} In molecular dynamics simulations or other calculations where configuration space coordinates are changed incrementally, the simplex that contained the previous simulation point would be a suitable candidate for $ \tilde{\mathcal{S}}_{\mathrm{init}}$. Unfortunately, this is not the case for good Monte Carlo simulations, which change the coordinates significantly in a single step of the random walk. As a result, we used k-trees\cite{Bentley:1975} for generating an approximation $ \mathbf{r}_{\mathrm{close}}$ for the mesh point closest to $\tilde{\mathbf{r}}$; once $ \mathbf{r}_{\mathrm{close}}$ is found, we randomly choose a simplex $ \tilde{\mathcal{S}}_{\mathrm{init}}$ that has $ \mathbf{r}_{\mathrm{close}}$ as its vertex. The computational cost of finding $ \mathbf{r}_{\mathrm{close}}$ scales logarithmically with the number of points in the mesh, and it's reasonable to assume that the computational cost of subsequent choosing of $ \tilde{\mathcal{S}}_{\mathrm{init}}$ and locating $\tilde{\mathcal{S}}$ is approximately constant for large enough numbers of mesh points. Since computational cost of simplex interpolation based on a known simplex does not depend on the number of mesh points, the total cost of our interpolation procedure scales logarithmically with the number of mesh points, as was mentioned in the Introduction.

\section{Numerical tests\label{sec:Applications}}

\subsection{Anharmonic oscillator\label{subsec:sqo_applications}}

\begin{figure}
[ptbh]\centering\includegraphics[width=\textwidth]{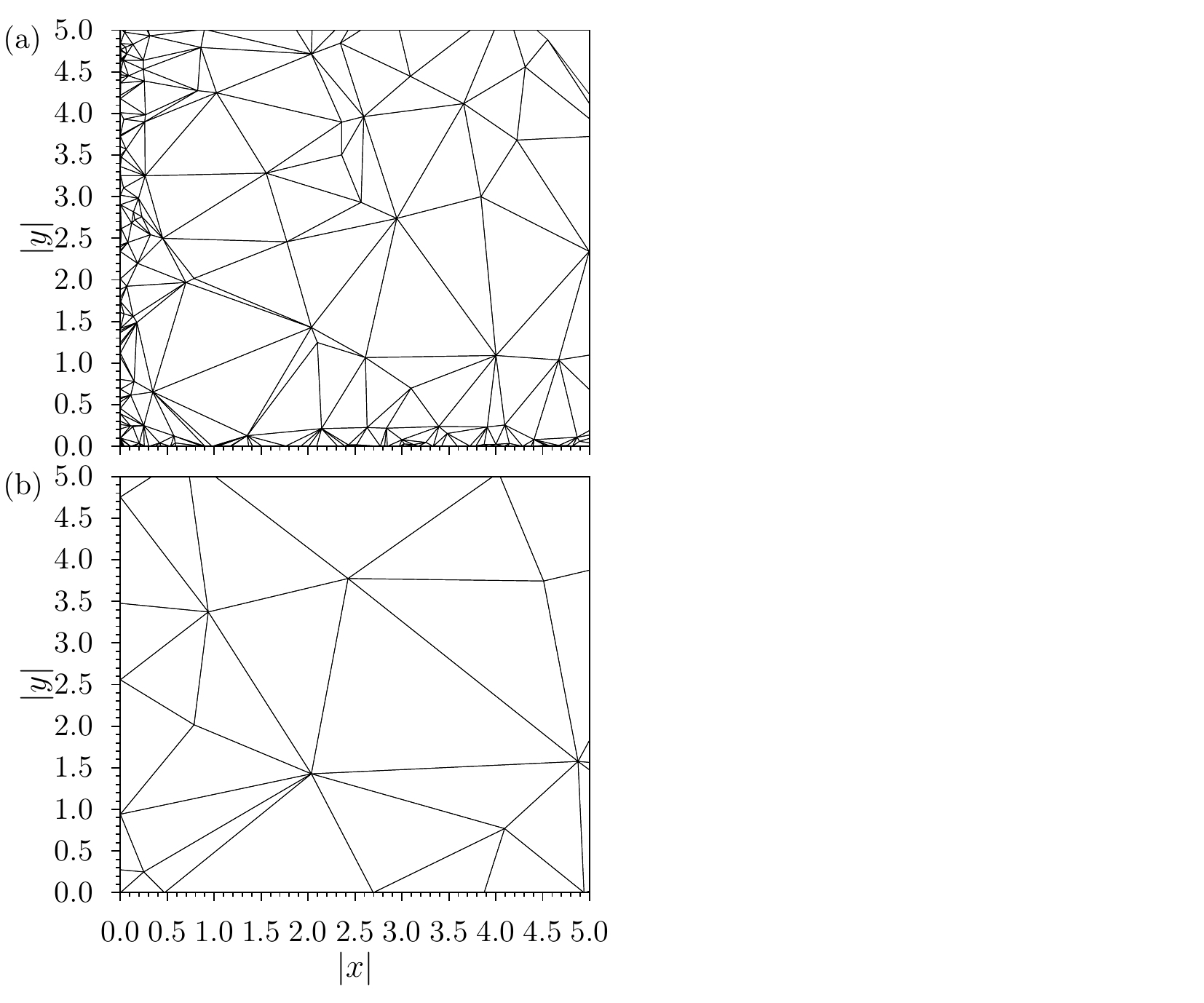}
\caption{\label{fig:simplices_outward_push}Mesh triangulations obtained in classical Monte Carlo simulations of a 2D harmonic oscillator without (a) and with (b) the ``outward push'' procedure.}
\end{figure}

\begin{figure}
[ptbh]\centering\includegraphics[width=\textwidth]{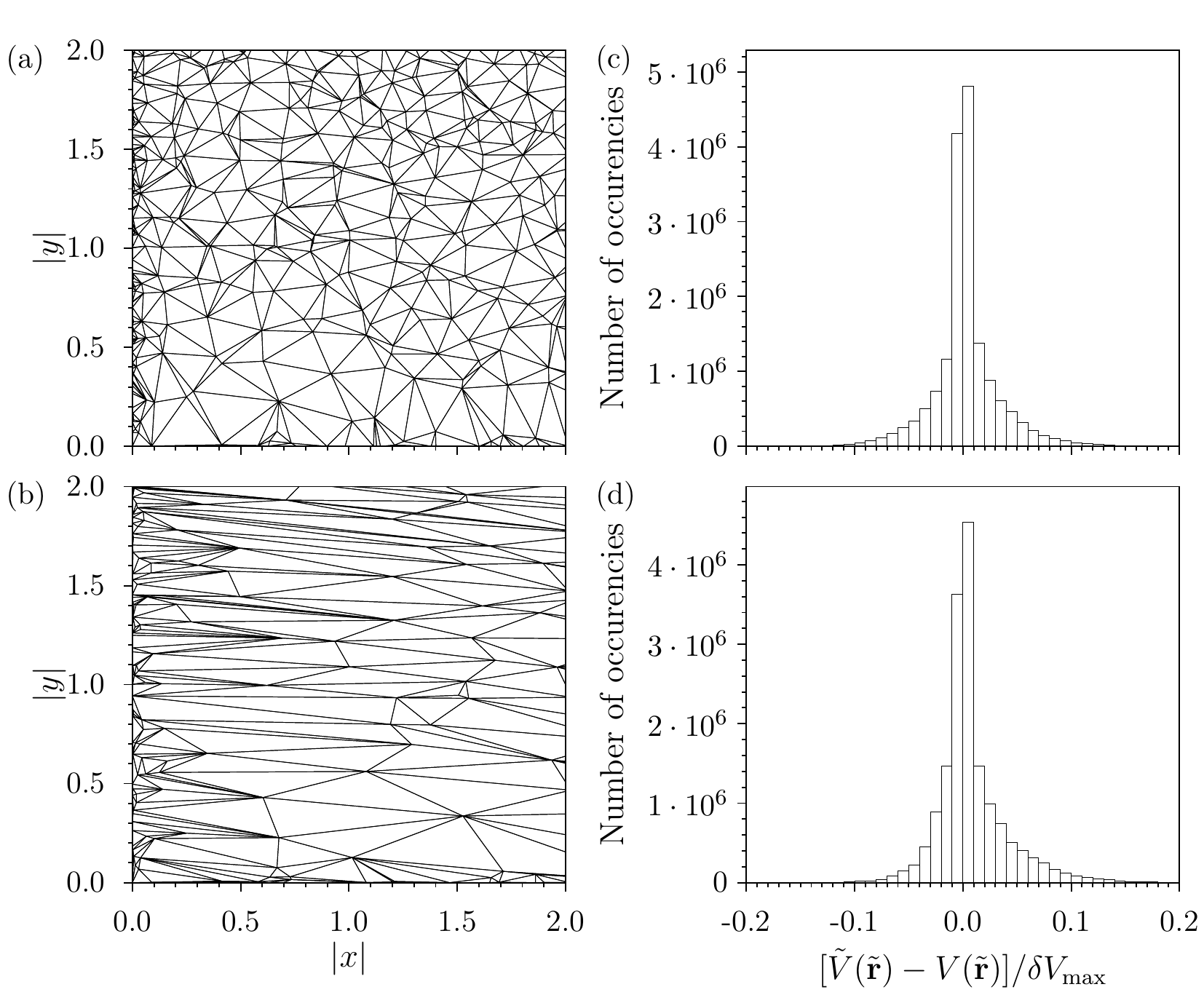}
\caption{\label{fig:simplices_non_Del}Mesh triangulations and distributions of the interpolation error obtained in classical Monte Carlo simulations of a 2D quartic oscillator~(\ref{eq:sqo_def}) with $ \epsilon_{\mathrm{anharm}}=0.01$ using the Delaunay [(a), (c)] or anisotropic [(b), (d)] triangulations.}
\end{figure}

\begin{figure}
[ptbh]\centering\includegraphics[width=\textwidth]{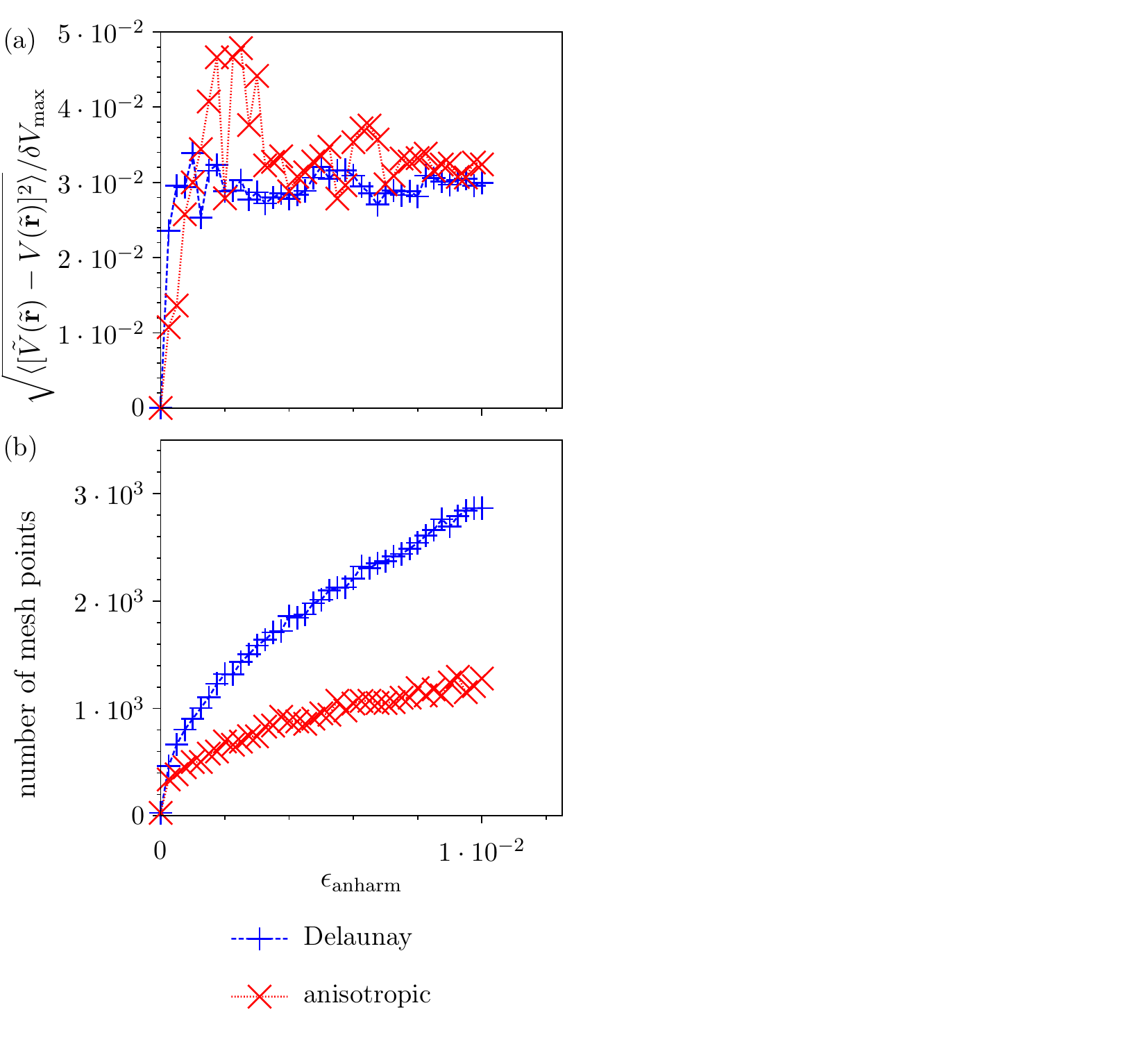}
\caption{\label{fig:error_nmesh_comp}Comparison of Delaunay and anisotropic triangulations applied in classical Monte Carlo simulations of two-dimensional quartic oscillators~(\ref{eq:sqo_def}) with different $ \epsilon_{\mathrm{anharm}}$.
(a) Root mean square errors (RMSEs) of interpolation $ \langle [\tilde{V}(\tilde{\mathbf{r}})-V(\tilde{\mathbf{r}})]^{2}\rangle^{1/2}$, (b) number of mesh points required.}
\end{figure}

A large number of molecular systems are close to harmonic in the most relevant part of their configuration space, so as a model problem we chose a system of two harmonic vibrational modes with a ``very anisotropic'' anharmonic perturbation,
\begin{equation}
 V(x, y)=x^{2}+y^{2}+\epsilon_{\mathrm{anharm}}[x^{4}+(4y)^{4}],
 \label{eq:sqo_def}
\end{equation}
where $ \epsilon_{\mathrm{anharm}}$ determines the ``anharmonicity'' of the potential. In all examples presented here we ran $ 2^{24}$ ($ \approx 1.68\cdot10^{7}$) step Monte Carlo simulations with inverse thermodynamic temperature $ \beta=1$ and $ \delta V_{\mathrm{max}}=3.125\cdot10^{-2}$; the mesh was constructed in $ |x|$ and $ |y|$ rather than $ x$ and $ y$ to capitalize on the potential's symmetry. The results are presented in Figs.~\ref{fig:simplices_outward_push}-\ref{fig:error_nmesh_comp}, with the RMSEs of interpolation and the number of points added to the mesh during simulations with $ \epsilon_{\mathrm{anharm}}=0$ and $ \epsilon_{\mathrm{anharm}}=0.01$ displayed in Table~\ref{tab:SQO_MC_errs}.

Figure~\ref{fig:simplices_outward_push} illustrates the reasoning behind introducing the ``outward push'' procedure (see Subsec.~\ref{subsec:mesh_update}) by comparing Delaunay triangulations generated at $ \epsilon_{\mathrm{anharm}}=0$ without [panel (a)] and with [panel (b)] the outward push. In this harmonic case the interpolation procedure is exact with a $\delta V(\tilde{\mathbf{r}})=0$. Comparing the two triangulations illustrates how introducing the ``outward push'' decreases the number of points used and prevents the algorithm from placing many mesh points close to $ |x|=0$ and $ |y|=0$ lines as $ \mathcal{C}_{\mathrm{mesh}}$ incrementally approaches them. As seen from Table~\ref{tab:SQO_MC_errs}, in this situation the ``outward push'' procedure lead almost to an order-of-magnitude decrease in the number of mesh points, even though, as illustrated by results for $ \epsilon_{\mathrm{anharm}}=0.01$ in Table~\ref{tab:SQO_MC_errs}, the improvement is definitely less drastic for anharmonic potentials where simplex size is also determined by the magnitude of $\delta V(\tilde{\mathbf{r}})$.

The motivation behind the anisotropic triangulation introduced in this work is illustrated by Fig.~\ref{fig:simplices_non_Del}, comparing the meshes and interpolation error distributions obtained using two different triangulations in Monte Carlo simulations for $ \epsilon_{\mathrm{anharm}}=0.01$. The simplices obtained with Delaunay and anisotropic triangulations are plotted in panels~(a) and~(b); switching to the anisotropic triangulation ``elongates'' triangles along $ |x|$ axis, as it should, judging by the form of anharmonic part of the potential~(\ref{eq:sqo_def}). While in this case the distribution of interpolation errors is not significantly affected, the number of mesh points added is decreased more than by a factor of two (see Table~\ref{tab:SQO_MC_errs}).

We also checked how the tendencies observed for $ \epsilon_{\mathrm{anharm}}=0.01$ hold for other values of $ \epsilon_{\mathrm{anharm}}$; Fig.~\ref{fig:error_nmesh_comp} demonstrates the resulting RMSEs of interpolation $\langle[\tilde{V}(\tilde{\mathbf{r}})-V(\tilde{\mathbf{r}})]^{2}\rangle^{1/2}$ [panel~(a)] and number of mesh points [panel~(b)]. Although changing the triangulation to anisotropic in this system increases slightly the interpolation errors,  the decrease in the number of mesh points is much more significant. Also note that the RMSE is always much smaller than $ \delta V_{\mathrm{max}}$, a tendency we observed for a wide range of potentials.

Lastly, recall that running the Monte Carlo simulations presented here involved $ 2^{24}+1\approx1.6\cdot10^{7}$ potential energy evaluations, and instead of exact calculations in each instance we used mere thousands of mesh points to reproduce these exact calculations with great precision (see Table~\ref{tab:SQO_MC_errs}). This demonstrates the potential of our method for speeding up practical calculations, a point elaborated further in the next subsection.

\begin{table}
\caption{Root mean square errors  (RMSEs) $ \langle [\tilde{V}(\tilde{\mathbf{r}})-V(\tilde{\mathbf{r}})]^{2}\rangle^{1/2}$ of interpolation and number of mesh points generated in Monte Carlo simulations of the quartic oscillator~(\ref{eq:sqo_def}) at $ \beta=1$ with different interpolation methods at values of $ \epsilon_{\mathrm{anharm}}$ used in Figs.~\ref{fig:simplices_outward_push} and~\ref{fig:simplices_non_Del}. The statistical errors of RMSEs of interpolation were estimated with block averaging.\cite{Flyvbjerg_Petersen:1989}\hfill}\label{tab:SQO_MC_errs}
\begin{ruledtabular}
\begin{tabular}{ccccccc}
triangulation & \multicolumn{4}{c}{Delaunay} & \multicolumn{2}{c}{anisotropic}\tabularnewline
\cline{2-5}
use of ``outward push'' & \multicolumn{2}{c}{no} & \multicolumn{2}{c}{yes} & \multicolumn{2}{c}{yes} \tabularnewline\hline
$ \epsilon_{\mathrm{anharm}}$ & \rot{RMSE $ \times10^{4}$ a.u.}& \rot{mesh points} & \rot{RMSE $ \times10^{4}$ a.u.} & \rot{mesh points} & \rot{RMSE $ \times10^{4}$ a.u.} & \rot{mesh points}\tabularnewline
\hline
$ 0\phantom{.01}$ & 0& 290 & 0& 28 & 0 & 28\tabularnewline
$ 0.01$ & $9.622\pm0.003$ & 2975
 & $9.350\pm0.003$   & 2864 & $ 10.120\pm0.004$&1278\tabularnewline
\end{tabular}
\end{ruledtabular}
\end{table}

\subsection{HCN/DCN equilibrium isotope effect}

In this subsection we combine our interpolation procedure with the path integral Monte Carlo method,\cite{Herman_Berne:1982,Tuckerman_Klein:1993} which accounts for nuclear quantum effects by replacing each atom of the simulated molecule with $ P$ replicas connected by harmonic forces.\cite{Feynman_Hibbs:1965} We combined the path integral Monte Carlo method with the free energy perturbation approach\cite{Perez_Lilienfeld:2011} (direct estimators\cite{Cheng_Ceriotti:2014}) for isotope fractionation to calculate the $\mathrm{HCN/DCN} $ equilibrium isotope effect defined as
\begin{equation}
 \mathrm{IE}=\frac{Q_\mathrm{DCN}}{Q_\mathrm{HCN}},
 \label{eq:IE_def}
\end{equation}
where $ Q$ denotes the partition function. The potential energy surface of $ \mathrm{HCN}$ was taken from Ref.~\onlinecite{Makhnev_Zobov:2018}. The interpolation algorithm used three internal coordinates that were defined in terms of atom radius-vectors $ \mathrm{r}_{\mathrm{D/H}}$, $ \mathrm{r}_{\mathrm{C}}$, and $ \mathrm{r}_{\mathrm{N}}$ as follows
\begin{align}
 x_{1}&=|\mathrm{r}_{\mathrm{C}}-\mathrm{r}_{\mathrm{N}}|,\\
 x_{2}&=\frac{(\mathrm{r}_{\mathrm{D/H}}-\mathrm{r}_{\mathrm{C}})\cdot(\mathrm{r}_{\mathrm{C}}-\mathrm{r}_{\mathrm{N}})}{x_{1}},\\
 x_{3}&=\sqrt{|\mathrm{r}_{\mathrm{D/H}}-\mathrm{r}_{\mathrm{C}}|^{2}-x_{2}^{2}}.
\end{align}
It is necessary to use the  ``outward push'' procedure to avoid the mesh approaching infinitely closely the $ x_{3}=0$ surface due to the $ x_{3}\geq0$ constraint, for reasons illustrated in Subsec.~\ref{subsec:sqo_applications}.

\subsubsection{Numerical details}
\label{subsubsec:HCN_num_det}

For each temperature $ T=200,300,\ldots,900,1000$ K we ran a path integral Monte Carlo simulation of $ \mathrm{DCN}$ with the isotope effects~(\ref{eq:IE_def}) calculated by averaging the corresponding mass-scaled direct isotope effect estimator. Each Monte Carlo simulation was of $ 1.25\cdot2^{23}$ ($ \approx 1.05\cdot10^{7}$) steps, with $ 20\%$ being displacements of the entire ring polymer as a whole and the other $ 80\%$ being staging transformation\cite{Sprik_Chandler:1985,Sprik_Chandler:1985_1} movements of one fourth of the ring-polymer. The first $ 20\%$ of the Monte Carlo simulations were discarded as a warmup, while during the rest of the simulation the mass-scaled direct estimator was calculated every $ 8$ Monte Carlo steps (to avoid wasting computational effort on calculating correlated samples); the statistical error of its average was estimated as the root mean square error evaluated with block averaging.\cite{Flyvbjerg_Petersen:1989} The number of replicas $ P $ were chosen as $ 256$ and $ 32$ for $ 200$ K and $ 1000$ K; it was verified with separate calculations that doubling $ P$ did not change the isotope effect by more than $ 1\%$. For the other temperatures the $ P $ was assigned by linear interpolation of $ P$ values as a function of $ 1/T$. We set $ \delta V_{\mathrm{max}}=10^{-4}$ a.u., and after each successful interpolation the algorithm had a $ 10^{-5}$ probability to carry out an additional exact potential energy calculation in order to estimate the RMSE of interpolation.

\subsubsection{Results and discussion}

In Table~\ref{tab:HCN_IE_vals}, isotope effects calculated with our interpolation algorithm are compared to benchmark values calculated with the original force field and with harmonic approximation\cite{Urey:1947,Wolfsberg_Rebelo:2010,Webb_Miller:2014} values. Interpolation allows reproducing benchmark isotope effect values with an error below $ 1\%$; the decent agreement between the harmonic approximation values and the formally exact path integral results are expected considering $ \mathrm{HCN}$ is a fairly harmonic molecule.

\begin{table}
\caption{$ \mathrm{HCN/DCN}$ isotope effect values calculated with the path integral Monte Carlo method and with the harmonic approximation. The path integral simulations were done using both the original force field  (the ``benchmark'' calculation) and our interpolation algorithm employing either the Delaunay or anisotropic triangulation. $ P$ is the number of imaginary-time slices used.\hfill}\label{tab:HCN_IE_vals}
\begin{ruledtabular}
\begin{tabular}{cccccc}
\multirow{2}{*}{$T$(K)} & \multirow{2}{*}{$P$} & \multicolumn{3}{c}{path integral calculations} & \multirow{2}{*}{\begin{tabular}{c}harmonic\tabularnewline approximation\end{tabular}}\tabularnewline
\cline{3-5}
& & Delaunay mesh & anisotropic mesh & benchmark & \tabularnewline
 \cline{1-6}
\pz200 & 256 & $ 4.5356\pm0.0019$ & $ 4.5379\pm0.0019$ & $ 4.5358\pm0.0018$ & 4.635 \tabularnewline
\pz300 & 162 & $ 3.1623 \pm0.0008$ & $ 3.1639\pm0.0009$ & $ 3.1623\pm0.0012$ & 3.228\tabularnewline
\pz400 & 116 & $ 2.5028\pm0.0008$& $ 2.5005\pm0.0009$& $ 2.5012\pm0.0009$ & 2.550\tabularnewline
\pz500 & \pz88 & $ 2.1178\pm0.0009$& $ 2.1166\pm0.0010$& $ 2.1190\pm0.0007$ &2.157\tabularnewline
\pz600 & \pz70 & $ 1.8684\pm0.0007$ & $ 1.8701\pm0.0007$& $ 1.8696\pm0.0007$ &1.903\tabularnewline
\pz700 & \pz56 & $ 1.6985\pm0.0006$& $ 1.6983\pm0.0005$ & $ 1.6985\pm0.0005$ &1.728\tabularnewline
\pz800 & \pz46 & $ 1.5746\pm0.0005$& $ 1.5751\pm0.0005$ & $1.5744\pm0.0005$ &1.600\tabularnewline
\pz900 & \pz38 & $ 1.4827\pm0.0005$& $ 1.4822\pm0.0005$ & $ 1.4824\pm0.0004$ &1.505\tabularnewline
1000 & \pz32 & $ 1.4113\pm0.0004$& $ 1.4115\pm0.0005$ & $ 1.4108\pm0.0004$ &1.431\tabularnewline
\end{tabular}
\end{ruledtabular}
\end{table}

The RMSEs of interpolation and the number of mesh points generated during the simulations are displayed in Table~\ref{tab:HCN_IE_errs}. If we used an expensive {\it ab initio} procedure for the exact potential, then the speedup due to the interpolation method would equal the ratio of the numbers of interpolated potential energy evaluations and the number of exact potential energy evaluations which approximately equals to the number of mesh points generated in the simulation. In this 3-dimensional problem this ratio is always of the order of $ 10^{4}$, indicating a large potential speedup. As discussed in Subsec.~\ref{subsec:interpolant}, we made an additional number of exact potential energy calculations to make sure that the choice of $ \delta V_{\mathrm{max}}$ guarantees an adequate interpolation accuracy, however the number of these additional calculations (approximately the number of potential evaluations during the calculation times $ 10^{-5}$, see Subsubsec.~\ref{subsubsec:HCN_num_det}) was always small compared to the number of mesh points, but still large enough to estimate the RMSE of interpolation with high precision. As was the case for the quartic oscillator, the mean square interpolation error is significantly smaller than $ |\delta V_{\mathrm{max}}|^{2}$. However, because $ \mathrm{HCN}$ is a very harmonic system, the anisotropic triangulation loses its advantage over the Delaunay triangulation. Both approaches behave similarly and, in fact, the anisotropic triangulation yields slightly higher interpolation errors and generates slightly more mesh points.

\begin{table}
\caption{Root mean square errors (RMSEs) $ \langle[\tilde{V}(\tilde{\mathbf{r}})-V(\tilde{\mathbf{r}})]^{2}\rangle^{1/2}$ of interpolation and number of mesh points generated for path integral $ \mathrm{HCN/DCN}$ isotope effect calculations along with the number of exact potential energy surface calculations that were required by the benchmark calculations. The statistical errors of RMSEs of interpolation were estimated with block averaging.\cite{Flyvbjerg_Petersen:1989}\hfill}\label{tab:HCN_IE_errs}
\begin{ruledtabular}
\begin{tabular}{cccccc}
\multirow{3}{*}{$ T $(K)}  & \multicolumn{4}{c}{Interpolation procedure} & \multirow{3}{*}{\begin{tabular}{c}number of exact\tabularnewline PES calculations in\tabularnewline benchmark simulation\end{tabular}}\tabularnewline
\cline{2-5}
& \multicolumn{2}{c}{Delaunay triangulation} & \multicolumn{2}{c}{anisotropic triangulation} & \tabularnewline\cline{2-3}\cline{4-5}
& RMSE $ \times 10^{6}$ a.u. & mesh points  & RMSE $ \times 10^{6}$ a.u. & mesh points &  \tabularnewline
\cline{1-6}
\pz200  & $3.13\pm0.04$& $ 3.11\cdot 10^{4}$  & $ 4.18\pm0.06$ & $3.15\cdot 10^{4}$ & $1.34\cdot 10^{9}$ \tabularnewline
\pz300  & $3.06\pm0.04$& $ 2.32\cdot 10^{4}$ & $4.56\pm0.17$ & $2.30\cdot 10^{4}$ & $ 8.45\cdot 10^{8}$ \tabularnewline
\pz400  & $2.98\pm0.05$& $ 2.07\cdot 10^{4}$  & $4.33\pm0.10$ & $1.84\cdot 10^{4}$ & $ 6.08\cdot 10^{8}$\tabularnewline
\pz500  & $3.09\pm0.06$& $ 1.96\cdot 10^{4}$  & $5.7\pz\pm0.8
\pz $ & $1.80\cdot 10^{4}$ & $ 4.61\cdot 10^{8}$ \tabularnewline
\pz600  & $3.04\pm0.07$& $ 1.95\cdot 10^{4}$ & $3.94\pm0.12$ & $1.90\cdot 10^{4}$ & $ 3.63\cdot 10^{8}$ \tabularnewline
\pz700  & $3.18\pm0.07$ & $ 1.96\cdot 10^{4}$ & $3.83\pm0.16$ & $2.65\cdot 10^{4}$ & $ 2.94\cdot 10^{8}$\tabularnewline
\pz800  & $2.94\pm0.08$ & $ 1.99\cdot 10^{4}$ & $5.4\pz\pm0.5\pz$ & $2.47\cdot 10^{4}$ & $ 2.37\cdot 10^{8}$\tabularnewline
\pz900  & $3.20\pm0.09$ & $ 2.00\cdot 10^{4}$  & $3.79\pm0.13
$ & $2.12\cdot 10^{4}$ & $ 1.95\cdot 10^{8}$\tabularnewline
1000 & $2.95\pm0.09$ & $ 2.09\cdot 10^{4}$  & $3.86\pm0.18$ & $2.47\cdot 10^{4}$ & $ 1.68\cdot 10^{8}$\tabularnewline
\end{tabular}
\end{ruledtabular}
\end{table}

We have also investigated how our method performs if the mesh created during one simulation is reused for simulations at other temperatures. One would expect that the best starting point would be the mesh generated during the highest temperature simulation, which—in accordance with classical Boltzmann distribution—tends to visit a larger region of configuration space during a fixed number of simulation steps. However, in our calculations of HCN isotope effect we found that using the lowest temperature was preferable.\footnote{Table~\ref{tab:HCN_IE_errs} shows that the largest number of mesh points was generated during the lowest temperature simulation. Because the spacing of mesh points does not depend on temperature, this suggests that the mesh generated at the lowest temperature covered the largest region of configuration space and, therefore, should be used as a starting mesh. This counterintuitive observation can be explained as follows: The lower the temperature, the greater the quantum delocalization of the ring polymer and the greater the explored region of configuration space. Although this delocalization diminishes at higher temperatures, it is eventually replaced by an ever increasing motion of the center of the ring polymer associated with the classical Boltzmann distribution. However, we cannot see this transition yet in Table~\ref{tab:HCN_IE_errs}, most likely because the quantum delocalization at lower temperatures is further increased due to the use mass-scaled direct estimators, which stretch the ring polymer by a factor of $\sqrt{2}$.\cite{Cheng_Ceriotti:2014}\protect\pseudodot.} The results are presented in Table~\ref{tab:HCN_IE_errs_reused_mesh}, which shows clearly that the number of extra mesh points that must be added to the mesh generated during the lowest temperature simulation is relatively small for simulations at all other temperatures.

\begin{table}
\caption{Root mean square errors (RMSEs) $ \langle[\tilde{V}(\tilde{\mathbf{r}})-V(\tilde{\mathbf{r}})]^{2}\rangle^{1/2}$ of interpolation and number of additional mesh points generated for path integral $ \mathrm{HCN/DCN}$ isotope effect calculations that reused the mesh generated during the calculations at $ 200$ K presented in Table~\ref{tab:HCN_IE_errs}. The statistical errors of RMSEs of interpolation were estimated with block averaging.\cite{Flyvbjerg_Petersen:1989}\hfill}\label{tab:HCN_IE_errs_reused_mesh}
\begin{ruledtabular}
\begin{tabular}{ccccc}
\multirow{4}{*}{$ T $(K)}  & \multicolumn{4}{c}{Interpolation procedure} \tabularnewline
\cline{2-5}
& \multicolumn{2}{c}{Delaunay triangulation} & \multicolumn{2}{c}{anisotropic triangulation} \tabularnewline\cline{2-3}\cline{4-5}
& \multirow{2}{*}{RMSE $ \times 10^{6}$ a.u.} & additional & \multirow{2}{*}{RMSE $ \times 10^{6}$ a.u.} & additional\tabularnewline
&  & mesh points  &  & mesh points\tabularnewline
\cline{1-5}
\pz200 & $3.13\pm0.04$& $\pz\pz\pz0$ & $ 4.18\pm0.06$ & $\pz\pz\pz0$\tabularnewline
\pz300 & $3.11\pm0.04$& $\pz503$ & $4.14\pm0.08$ & $\pz786$\tabularnewline
\pz400 & $3.09\pm0.05$& $\pz413$ & $4.15\pm0.09$ & $\pz713$\tabularnewline
\pz500 & $3.09\pm0.06$& $\pz747$ & $3.96\pm0.09$ & $\pz803$\tabularnewline
\pz600 & $2.97\pm0.06$& $\pz774$ & $3.94\pm0.10$ & $1102$\tabularnewline
\pz700 & $3.15\pm0.07$ & $1591$ & $4.08\pm0.15$ & $1635$\tabularnewline
\pz800 & $3.03\pm0.08$ & $2421$ & $3.81\pm0.12$ & $2467$ \tabularnewline
\pz900 & $3.20\pm0.09$ & $2795$  & $3.87\pm0.15$ & $2357$\tabularnewline
1000 & $2.94\pm0.08$ & $3048$  & $4.02\pm0.19$ & $3235$\tabularnewline
\end{tabular}
\end{ruledtabular}
\end{table}

\section{Conclusion}

We have proposed an algorithm for interpolating potential energy values from the values of the potential energy and its gradient calculated and stored for points of a mesh generated during a Monte Carlo simulation. The interpolation procedure is exact in harmonic systems, while in anharmonic systems its accuracy depends on the triangulation procedure chosen for the mesh. For the latter, we considered two choices: the previously used Delaunay triangulation and an anisotropic triangulation designed to decrease the interpolation error. Both triangulations combined with subsequent interpolation resulted in a very large reduction of potential energy evaluations in comparison with a purely on-the-fly approach. Moreover, we found that for nearly harmonic systems the two triangulations give similar results, with Delaunay triangulation demonstrating superior performance in some cases, but for more anharmonic systems the proposed anisotropic triangulation achieves similar interpolation errors with significantly fewer mesh points. The {\it ad hoc} procedure used for construction of such anisotropic triangulations may be used to improve performance of other interpolants,\cite{Alfeld:1984,Alfeld:1985,Guessab_Schmeisser:2006,Dell'Accio_Zerroudi:2018} even though a different definition of $ \delta V(\tilde{\mathbf{r}})$~[Eq.~(\ref{eq:delta_V_definition})] may prove more convenient.

To combine our interpolation algorithm with classical or semiclassical molecular dynamics simulations, one may need to use a different interpolant, as mentioned in Subsec.~\ref{subsec:interpolant}; the ``outward push'' procedure would also need to be extended to points added inside $ \mathcal{C}_{\mathrm{mesh}}$ to avoid forming nearly degenerate simplices. By contrast, as mentioned in Subsec.~\ref{subsec:simplex_search}, searching for the simplex used in interpolation should become even simpler.

It is important to discuss the scaling of our interpolation procedure with respect to two parameters: number of points in the mesh and dimensionality. For the former, adding new points to the mesh is done at a cost that does not depend on the number of points already in the mesh and the cost of finding the simplex used in interpolation scales logarithmically with the number of mesh points; calculating the interpolant costs the same regardless of the number of mesh points. This behavior compares favourably to Shepard interpolation and Gaussian process regression, which utilize functions whose evaluation cost is proportional to the number of mesh points; triangulating the mesh is a natural way to avoid the issue. However, the cost of storing and updating the triangulation increases dramatically with dimensionality\cite{Hornus_Boissonnat:2008} even if one does not take into account the increase in the needed number of mesh points (which also grows quickly with dimensionality, at least for lower dimensions). This problem is likely to be decisive if one wanted to apply our method to systems of dimensionality six and higher (corresponding to molecules with four atoms and more). Yet, in this work we demonstrated a significant potential speedup achieved by such algorithms in the simulations of two- and three-dimensional systems.

\begin{acknowledgments}
The authors acknowledge the financial support from  the European Research Council (ERC) under the European Union's Horizon 2020 research and innovation programme (grant agreement No. 683069 -- MOLEQULE). We also thank Fabio Albertani for useful discussions.
\end{acknowledgments}

\bibliographystyle{aipnum4-1}

\begin{thebibliography}{57}%
\makeatletter
\providecommand \@ifxundefined [1]{%
 \@ifx{#1\undefined}
}%
\providecommand \@ifnum [1]{%
 \ifnum #1\expandafter \@firstoftwo
 \else \expandafter \@secondoftwo
 \fi
}%
\providecommand \@ifx [1]{%
 \ifx #1\expandafter \@firstoftwo
 \else \expandafter \@secondoftwo
 \fi
}%
\providecommand \natexlab [1]{#1}%
\providecommand \enquote  [1]{``#1''}%
\providecommand \bibnamefont  [1]{#1}%
\providecommand \bibfnamefont [1]{#1}%
\providecommand \citenamefont [1]{#1}%
\providecommand \href@noop [0]{\@secondoftwo}%
\providecommand \href [0]{\begingroup \@sanitize@url \@href}%
\providecommand \@href[1]{\@@startlink{#1}\@@href}%
\providecommand \@@href[1]{\endgroup#1\@@endlink}%
\providecommand \@sanitize@url [0]{\catcode `\\12\catcode `\$12\catcode
  `\&12\catcode `\#12\catcode `\^12\catcode `\_12\catcode `\%12\relax}%
\providecommand \@@startlink[1]{}%
\providecommand \@@endlink[0]{}%
\providecommand \url  [0]{\begingroup\@sanitize@url \@url }%
\providecommand \@url [1]{\endgroup\@href {#1}{\urlprefix }}%
\providecommand \urlprefix  [0]{URL }%
\providecommand \Eprint [0]{\href }%
\providecommand \doibase [0]{http://dx.doi.org/}%
\providecommand \selectlanguage [0]{\@gobble}%
\providecommand \bibinfo  [0]{\@secondoftwo}%
\providecommand \bibfield  [0]{\@secondoftwo}%
\providecommand \translation [1]{[#1]}%
\providecommand \BibitemOpen [0]{}%
\providecommand \bibitemStop [0]{}%
\providecommand \bibitemNoStop [0]{.\EOS\space}%
\providecommand \EOS [0]{\spacefactor3000\relax}%
\providecommand \BibitemShut  [1]{\csname bibitem#1\endcsname}%
\let\auto@bib@innerbib\@empty
\bibitem [{\citenamefont {Song}\ and\ \citenamefont
  {Mart\'{i}nez}(2016)}]{Song_Martinez:2016}%
  \BibitemOpen
  \bibfield  {author} {\bibinfo {author} {\bibfnamefont {C.}~\bibnamefont
  {Song}}\ and\ \bibinfo {author} {\bibfnamefont {T.~J.}\ \bibnamefont
  {Mart\'{i}nez}},\ }\href {https://doi.org/10.1063/1.4948438} {\bibfield
  {journal} {\bibinfo  {journal} {J. Chem. Phys.}\ }\textbf {\bibinfo {volume}
  {144}},\ \bibinfo {eid} {174111} (\bibinfo {year} {2016})}\BibitemShut
  {NoStop}%
\bibitem [{\citenamefont {Jeanmairet}, \citenamefont {Sharma},\ and\
  \citenamefont {Alavi}(2017)}]{Jeanmairet_Alavi:2017}%
  \BibitemOpen
  \bibfield  {author} {\bibinfo {author} {\bibfnamefont {G.}~\bibnamefont
  {Jeanmairet}}, \bibinfo {author} {\bibfnamefont {S.}~\bibnamefont {Sharma}},
  \ and\ \bibinfo {author} {\bibfnamefont {A.}~\bibnamefont {Alavi}},\ }\href
  {https://doi.org/10.1063/1.4974177} {\bibfield  {journal} {\bibinfo
  {journal} {J. Chem. Phys.}\ }\textbf {\bibinfo {volume} {146}},\ \bibinfo
  {eid} {044107} (\bibinfo {year} {2017})}\BibitemShut {NoStop}%
\bibitem [{\citenamefont {P\'{e}rez}\ \emph {et~al.}(2010)\citenamefont
  {P\'{e}rez}, \citenamefont {Tuckerman}, \citenamefont {Hjalmarson},\ and\
  \citenamefont {von Lilienfeld}}]{Perez_Lilienfeld:2010}%
  \BibitemOpen
  \bibfield  {author} {\bibinfo {author} {\bibfnamefont {A.}~\bibnamefont
  {P\'{e}rez}}, \bibinfo {author} {\bibfnamefont {M.~E.}\ \bibnamefont
  {Tuckerman}}, \bibinfo {author} {\bibfnamefont {H.~P.}\ \bibnamefont
  {Hjalmarson}}, \ and\ \bibinfo {author} {\bibfnamefont {O.~A.}\ \bibnamefont
  {von Lilienfeld}},\ }\href {https://doi.org/10.1021/ja102004b} {\bibfield
  {journal} {\bibinfo  {journal} {J. Am. Chem. Soc.}\ }\textbf {\bibinfo
  {volume} {132}},\ \bibinfo {eid} {11510} (\bibinfo {year}
  {2010})}\BibitemShut {NoStop}%
\bibitem [{\citenamefont {Gasparotto}, \citenamefont {Hassanali},\ and\
  \citenamefont {Ceriotti}(2016)}]{Gasparotto_Ceriotti:2016}%
  \BibitemOpen
  \bibfield  {author} {\bibinfo {author} {\bibfnamefont {P.}~\bibnamefont
  {Gasparotto}}, \bibinfo {author} {\bibfnamefont {A.~A.}\ \bibnamefont
  {Hassanali}}, \ and\ \bibinfo {author} {\bibfnamefont {M.}~\bibnamefont
  {Ceriotti}},\ }\href {https://doi.org/10.1021/acs.jctc.5b01138} {\bibfield
  {journal} {\bibinfo  {journal} {J. Chem. Theory Comput.}\ }\textbf {\bibinfo
  {volume} {12}},\ \bibinfo {eid} {1953} (\bibinfo {year} {2016})}\BibitemShut
  {NoStop}%
\bibitem [{\citenamefont {de~Tudela}\ and\ \citenamefont
  {Marx}(2017)}]{PerezDeTudela_Marx:2017}%
  \BibitemOpen
  \bibfield  {author} {\bibinfo {author} {\bibfnamefont {R.~P.}\ \bibnamefont
  {de~Tudela}}\ and\ \bibinfo {author} {\bibfnamefont {D.}~\bibnamefont
  {Marx}},\ }\href {https://doi.org/10.1103/PhysRevLett.119.223001} {\bibfield
  {journal} {\bibinfo  {journal} {Phys. Rev. Lett.}\ }\textbf {\bibinfo
  {volume} {119}},\ \bibinfo {eid} {223001} (\bibinfo {year}
  {2017})}\BibitemShut {NoStop}%
\bibitem [{\citenamefont {Worth}, \citenamefont {Robb},\ and\ \citenamefont
  {Burghardt}(2004)}]{Worth_Burghardt:2004}%
  \BibitemOpen
  \bibfield  {author} {\bibinfo {author} {\bibfnamefont {G.~A.}\ \bibnamefont
  {Worth}}, \bibinfo {author} {\bibfnamefont {M.~A.}\ \bibnamefont {Robb}}, \
  and\ \bibinfo {author} {\bibfnamefont {I.}~\bibnamefont {Burghardt}},\ }\href
  {https://doi.org/10.1039/B314253A} {\bibfield  {journal} {\bibinfo  {journal}
  {Faraday Discuss.}\ }\textbf {\bibinfo {volume} {127}},\ \bibinfo {eid} {307}
  (\bibinfo {year} {2004})}\BibitemShut {NoStop}%
\bibitem [{\citenamefont {Ianconescu}, \citenamefont {Tatchen},\ and\
  \citenamefont {Pollak}(2013)}]{Ianconescu_Pollak:2013}%
  \BibitemOpen
  \bibfield  {author} {\bibinfo {author} {\bibfnamefont {R.}~\bibnamefont
  {Ianconescu}}, \bibinfo {author} {\bibfnamefont {J.}~\bibnamefont {Tatchen}},
  \ and\ \bibinfo {author} {\bibfnamefont {E.}~\bibnamefont {Pollak}},\ }\href
  {https://doi.org/10.1063/1.4825040} {\bibfield  {journal} {\bibinfo
  {journal} {J. Chem. Phys.}\ }\textbf {\bibinfo {volume} {139}},\ \bibinfo
  {eid} {154311} (\bibinfo {year} {2013})}\BibitemShut {NoStop}%
\bibitem [{\citenamefont {Curchod}\ and\ \citenamefont
  {Mart\'{i}nez}(2018)}]{Curchod_Martinez:2018}%
  \BibitemOpen
  \bibfield  {author} {\bibinfo {author} {\bibfnamefont {B.~F.~E.}\
  \bibnamefont {Curchod}}\ and\ \bibinfo {author} {\bibfnamefont {T.~J.}\
  \bibnamefont {Mart\'{i}nez}},\ }\href
  {https://doi.org/10.1021/acs.chemrev.7b00423} {\bibfield  {journal} {\bibinfo
   {journal} {Chem. Rev.}\ }\textbf {\bibinfo {volume} {118}},\ \bibinfo {eid}
  {3305} (\bibinfo {year} {2018})}\BibitemShut {NoStop}%
\bibitem [{\citenamefont {Spinlove}\ \emph {et~al.}(2018)\citenamefont
  {Spinlove}, \citenamefont {Richings}, \citenamefont {Robb},\ and\
  \citenamefont {Worth}}]{Spinlove_Worth:2018}%
  \BibitemOpen
  \bibfield  {author} {\bibinfo {author} {\bibfnamefont {K.~E.}\ \bibnamefont
  {Spinlove}}, \bibinfo {author} {\bibfnamefont {G.~W.}\ \bibnamefont
  {Richings}}, \bibinfo {author} {\bibfnamefont {M.~A.}\ \bibnamefont {Robb}},
  \ and\ \bibinfo {author} {\bibfnamefont {G.~A.}\ \bibnamefont {Worth}},\
  }\href {https://doi.org/10.1039/C8FD00090E} {\bibfield  {journal} {\bibinfo
  {journal} {Faraday Discuss.}\ }\textbf {\bibinfo {volume} {212}},\ \bibinfo
  {eid} {191} (\bibinfo {year} {2018})}\BibitemShut {NoStop}%
\bibitem [{\citenamefont {Patoz}, \citenamefont {Begu\v{s}i\'{c}},\ and\
  \citenamefont {Van\'{i}\v{c}ek}(2018)}]{Patoz_Vanicek:2018}%
  \BibitemOpen
  \bibfield  {author} {\bibinfo {author} {\bibfnamefont {A.}~\bibnamefont
  {Patoz}}, \bibinfo {author} {\bibfnamefont {T.}~\bibnamefont
  {Begu\v{s}i\'{c}}}, \ and\ \bibinfo {author} {\bibfnamefont {J.}~\bibnamefont
  {Van\'{i}\v{c}ek}},\ }\href {https://doi.org/10.1021/acs.jpclett.8b00827}
  {\bibfield  {journal} {\bibinfo  {journal} {J. Phys. Chem. Lett.}\ }\textbf
  {\bibinfo {volume} {9}},\ \bibinfo {eid} {2367} (\bibinfo {year}
  {2018})}\BibitemShut {NoStop}%
\bibitem [{\citenamefont {Micciarelli}\ \emph {et~al.}(2019)\citenamefont
  {Micciarelli}, \citenamefont {Gabas}, \citenamefont {Conte},\ and\
  \citenamefont {Ceotto}}]{Micciarelli_Ceotto:2019}%
  \BibitemOpen
  \bibfield  {author} {\bibinfo {author} {\bibfnamefont {M.}~\bibnamefont
  {Micciarelli}}, \bibinfo {author} {\bibfnamefont {F.}~\bibnamefont {Gabas}},
  \bibinfo {author} {\bibfnamefont {R.}~\bibnamefont {Conte}}, \ and\ \bibinfo
  {author} {\bibfnamefont {M.}~\bibnamefont {Ceotto}},\ }\href
  {https://doi.org/10.1063/1.5096968} {\bibfield  {journal} {\bibinfo
  {journal} {J. Chem. Phys.}\ }\textbf {\bibinfo {volume} {150}},\ \bibinfo
  {eid} {184113} (\bibinfo {year} {2019})}\BibitemShut {NoStop}%
\bibitem [{\citenamefont {Laude}\ \emph {et~al.}(2018)\citenamefont {Laude},
  \citenamefont {Calderini}, \citenamefont {Tew},\ and\ \citenamefont
  {Richardson}}]{Laude_Richardson:2018}%
  \BibitemOpen
  \bibfield  {author} {\bibinfo {author} {\bibfnamefont {G.}~\bibnamefont
  {Laude}}, \bibinfo {author} {\bibfnamefont {D.}~\bibnamefont {Calderini}},
  \bibinfo {author} {\bibfnamefont {D.~P.}\ \bibnamefont {Tew}}, \ and\
  \bibinfo {author} {\bibfnamefont {J.~O.}\ \bibnamefont {Richardson}},\ }\href
  {https://doi.org/10.1039/C8FD00085A} {\bibfield  {journal} {\bibinfo
  {journal} {Faraday Discuss.}\ }\textbf {\bibinfo {volume} {212}},\ \bibinfo
  {eid} {237} (\bibinfo {year} {2018})}\BibitemShut {NoStop}%
\bibitem [{\citenamefont {Conte}\ \emph {et~al.}(2019)\citenamefont {Conte},
  \citenamefont {Gabas}, \citenamefont {Botti}, \citenamefont {Zhuang},\ and\
  \citenamefont {Ceotto}}]{Conte_Ceotto:2019}%
  \BibitemOpen
  \bibfield  {author} {\bibinfo {author} {\bibfnamefont {R.}~\bibnamefont
  {Conte}}, \bibinfo {author} {\bibfnamefont {F.}~\bibnamefont {Gabas}},
  \bibinfo {author} {\bibfnamefont {G.}~\bibnamefont {Botti}}, \bibinfo
  {author} {\bibfnamefont {Y.}~\bibnamefont {Zhuang}}, \ and\ \bibinfo {author}
  {\bibfnamefont {M.}~\bibnamefont {Ceotto}},\ }\href
  {https://doi.org/10.1063/1.5109086} {\bibfield  {journal} {\bibinfo
  {journal} {J. Chem. Phys.}\ }\textbf {\bibinfo {volume} {150}},\ \bibinfo
  {eid} {244118} (\bibinfo {year} {2019})}\BibitemShut {NoStop}%
\bibitem [{\citenamefont {Ischtwan}\ and\ \citenamefont
  {Collins}(1994)}]{Ischtwan_Collins:1994}%
  \BibitemOpen
  \bibfield  {author} {\bibinfo {author} {\bibfnamefont {J.}~\bibnamefont
  {Ischtwan}}\ and\ \bibinfo {author} {\bibfnamefont {M.~A.}\ \bibnamefont
  {Collins}},\ }\href {https://doi.org/10.1063/1.466801} {\bibfield  {journal}
  {\bibinfo  {journal} {J. Chem. Phys.}\ }\textbf {\bibinfo {volume} {100}},\
  \bibinfo {eid} {8080} (\bibinfo {year} {1994})}\BibitemShut {NoStop}%
\bibitem [{\citenamefont {Frankcombe}, \citenamefont {Collins},\ and\
  \citenamefont {Worth}(2010)}]{Frankcombe_Worth:2010}%
  \BibitemOpen
  \bibfield  {author} {\bibinfo {author} {\bibfnamefont {T.~J.}\ \bibnamefont
  {Frankcombe}}, \bibinfo {author} {\bibfnamefont {M.~A.}\ \bibnamefont
  {Collins}}, \ and\ \bibinfo {author} {\bibfnamefont {G.~A.}\ \bibnamefont
  {Worth}},\ }\href {https://doi.org/10.1016/j.cplett.2010.02.068} {\bibfield
  {journal} {\bibinfo  {journal} {Chem. Phys. Lett.}\ }\textbf {\bibinfo
  {volume} {489}},\ \bibinfo {eid} {242} (\bibinfo {year} {2010})}\BibitemShut
  {NoStop}%
\bibitem [{\citenamefont {Evenhuis}\ and\ \citenamefont
  {Mart{\'i}nez}(2011)}]{Evenhuis_Martinez:2011}%
  \BibitemOpen
  \bibfield  {author} {\bibinfo {author} {\bibfnamefont {C.}~\bibnamefont
  {Evenhuis}}\ and\ \bibinfo {author} {\bibfnamefont {T.~J.}\ \bibnamefont
  {Mart{\'i}nez}},\ }\href {https://doi.org/10.1063/1.3660686} {\bibfield
  {journal} {\bibinfo  {journal} {J. Chem. Phys.}\ }\textbf {\bibinfo {volume}
  {135}},\ \bibinfo {eid} {224110} (\bibinfo {year} {2011})}\BibitemShut
  {NoStop}%
\bibitem [{\citenamefont {Kim}\ and\ \citenamefont
  {Rhee}(2016)}]{Kim_Rhee:2016}%
  \BibitemOpen
  \bibfield  {author} {\bibinfo {author} {\bibfnamefont {C.~W.}\ \bibnamefont
  {Kim}}\ and\ \bibinfo {author} {\bibfnamefont {Y.~M.}\ \bibnamefont {Rhee}},\
  }\href {https://doi.org/10.1021/acs.jctc.6b00647} {\bibfield  {journal}
  {\bibinfo  {journal} {J. Chem. Theory Comput.}\ }\textbf {\bibinfo {volume}
  {12}},\ \bibinfo {eid} {5235} (\bibinfo {year} {2016})}\BibitemShut {NoStop}%
\bibitem [{\citenamefont {Huang}, \citenamefont {Braams},\ and\ \citenamefont
  {Bowman}(2005)}]{Huang_Bowman:2005}%
  \BibitemOpen
  \bibfield  {author} {\bibinfo {author} {\bibfnamefont {X.}~\bibnamefont
  {Huang}}, \bibinfo {author} {\bibfnamefont {B.~J.}\ \bibnamefont {Braams}}, \
  and\ \bibinfo {author} {\bibfnamefont {J.~M.}\ \bibnamefont {Bowman}},\
  }\href {https://doi.org/10.1063/1.1834500} {\bibfield  {journal} {\bibinfo
  {journal} {J. Chem. Phys.}\ }\textbf {\bibinfo {volume} {122}},\ \bibinfo
  {eid} {044308} (\bibinfo {year} {2005})}\BibitemShut {NoStop}%
\bibitem [{\citenamefont {Guo}\ \emph {et~al.}(2004)\citenamefont {Guo},
  \citenamefont {Kawano}, \citenamefont {Thompson}, \citenamefont {Wagner},\
  and\ \citenamefont {Minkoff}}]{Guo_Minkoff:2004}%
  \BibitemOpen
  \bibfield  {author} {\bibinfo {author} {\bibfnamefont {Y.}~\bibnamefont
  {Guo}}, \bibinfo {author} {\bibfnamefont {A.}~\bibnamefont {Kawano}},
  \bibinfo {author} {\bibfnamefont {D.~L.}\ \bibnamefont {Thompson}}, \bibinfo
  {author} {\bibfnamefont {A.~F.}\ \bibnamefont {Wagner}}, \ and\ \bibinfo
  {author} {\bibfnamefont {M.}~\bibnamefont {Minkoff}},\ }\href
  {https://doi.org/10.1063/1.1777572} {\bibfield  {journal} {\bibinfo
  {journal} {J. Chem. Phys.}\ }\textbf {\bibinfo {volume} {121}},\ \bibinfo
  {eid} {5091} (\bibinfo {year} {2004})}\BibitemShut {NoStop}%
\bibitem [{\citenamefont {Dawes}\ \emph {et~al.}(2010)\citenamefont {Dawes},
  \citenamefont {Wang}, \citenamefont {Jasper},\ and\ \citenamefont {{T.
  Carrington Jr.}}}]{Dawes_Carrington:2010}%
  \BibitemOpen
  \bibfield  {author} {\bibinfo {author} {\bibfnamefont {R.}~\bibnamefont
  {Dawes}}, \bibinfo {author} {\bibfnamefont {X.-G.}\ \bibnamefont {Wang}},
  \bibinfo {author} {\bibfnamefont {A.~W.}\ \bibnamefont {Jasper}}, \ and\
  \bibinfo {author} {\bibnamefont {{T. Carrington Jr.}}},\ }\href
  {https://doi.org/10.1063/1.3494542} {\bibfield  {journal} {\bibinfo
  {journal} {J. Chem. Phys.}\ }\textbf {\bibinfo {volume} {133}},\ \bibinfo
  {eid} {134304} (\bibinfo {year} {2010})}\BibitemShut {NoStop}%
\bibitem [{\citenamefont {Alborzpour}, \citenamefont {Tew},\ and\ \citenamefont
  {Habershon}(2016)}]{Alborzpour_Habershon:2016}%
  \BibitemOpen
  \bibfield  {author} {\bibinfo {author} {\bibfnamefont {J.~P.}\ \bibnamefont
  {Alborzpour}}, \bibinfo {author} {\bibfnamefont {D.~P.}\ \bibnamefont {Tew}},
  \ and\ \bibinfo {author} {\bibfnamefont {S.}~\bibnamefont {Habershon}},\
  }\href {https://doi.org/10.1063/1.4964902} {\bibfield  {journal} {\bibinfo
  {journal} {J. Chem. Phys.}\ }\textbf {\bibinfo {volume} {145}},\ \bibinfo
  {eid} {174112} (\bibinfo {year} {2016})}\BibitemShut {NoStop}%
\bibitem [{\citenamefont {Richings}\ and\ \citenamefont
  {Habershon}(2017)}]{Richings_Habershon:2017}%
  \BibitemOpen
  \bibfield  {author} {\bibinfo {author} {\bibfnamefont {G.~W.}\ \bibnamefont
  {Richings}}\ and\ \bibinfo {author} {\bibfnamefont {S.}~\bibnamefont
  {Habershon}},\ }\href {https://doi.org/10.1016/j.cplett.2017.01.063}
  {\bibfield  {journal} {\bibinfo  {journal} {Chem. Phys. Lett.}\ }\textbf
  {\bibinfo {volume} {683}},\ \bibinfo {eid} {228} (\bibinfo {year}
  {2017})}\BibitemShut {NoStop}%
\bibitem [{\citenamefont {Blank}\ \emph {et~al.}(1995)\citenamefont {Blank},
  \citenamefont {Brown}, \citenamefont {Calhoun},\ and\ \citenamefont
  {Doren}}]{Blank_Doren:1995}%
  \BibitemOpen
  \bibfield  {author} {\bibinfo {author} {\bibfnamefont {T.~B.}\ \bibnamefont
  {Blank}}, \bibinfo {author} {\bibfnamefont {S.~D.}\ \bibnamefont {Brown}},
  \bibinfo {author} {\bibfnamefont {A.~W.}\ \bibnamefont {Calhoun}}, \ and\
  \bibinfo {author} {\bibfnamefont {D.~J.}\ \bibnamefont {Doren}},\ }\href
  {https://doi.org/10.1063/1.469597} {\bibfield  {journal} {\bibinfo  {journal}
  {J. Chem. Phys.}\ }\textbf {\bibinfo {volume} {103}},\ \bibinfo {eid} {4129}
  (\bibinfo {year} {1995})}\BibitemShut {NoStop}%
\bibitem [{\citenamefont {Manzhos}\ \emph {et~al.}(2006)\citenamefont
  {Manzhos}, \citenamefont {Wang}, \citenamefont {Dawes},\ and\ \citenamefont
  {{T. Carrington Jr.}}}]{Manzhos_Carrington:2006}%
  \BibitemOpen
  \bibfield  {author} {\bibinfo {author} {\bibfnamefont {S.}~\bibnamefont
  {Manzhos}}, \bibinfo {author} {\bibfnamefont {X.}~\bibnamefont {Wang}},
  \bibinfo {author} {\bibfnamefont {R.}~\bibnamefont {Dawes}}, \ and\ \bibinfo
  {author} {\bibnamefont {{T. Carrington Jr.}}},\ }\href
  {https://doi.org/10.1021/jp055253z} {\bibfield  {journal} {\bibinfo
  {journal} {J. Phys. Chem. A}\ }\textbf {\bibinfo {volume} {110}},\ \bibinfo
  {eid} {5295} (\bibinfo {year} {2006})}\BibitemShut {NoStop}%
\bibitem [{\citenamefont {Malshe}\ \emph {et~al.}(2010)\citenamefont {Malshe},
  \citenamefont {Raff}, \citenamefont {Hagan}, \citenamefont {Bukkapatnam},\
  and\ \citenamefont {Komanduri}}]{Malshe_Komanduri:2010}%
  \BibitemOpen
  \bibfield  {author} {\bibinfo {author} {\bibfnamefont {M.}~\bibnamefont
  {Malshe}}, \bibinfo {author} {\bibfnamefont {L.~M.}\ \bibnamefont {Raff}},
  \bibinfo {author} {\bibfnamefont {M.}~\bibnamefont {Hagan}}, \bibinfo
  {author} {\bibfnamefont {S.}~\bibnamefont {Bukkapatnam}}, \ and\ \bibinfo
  {author} {\bibfnamefont {R.}~\bibnamefont {Komanduri}},\ }\href
  {https://doi.org/10.1063/1.3431624} {\bibfield  {journal} {\bibinfo
  {journal} {J. Chem. Phys.}\ }\textbf {\bibinfo {volume} {132}},\ \bibinfo
  {eid} {204103} (\bibinfo {year} {2010})}\BibitemShut {NoStop}%
\bibitem [{\citenamefont {Gastegger}, \citenamefont {Behler},\ and\
  \citenamefont {Marquetand}(2017)}]{Gastegger_Marquetand:2017}%
  \BibitemOpen
  \bibfield  {author} {\bibinfo {author} {\bibfnamefont {M.}~\bibnamefont
  {Gastegger}}, \bibinfo {author} {\bibfnamefont {J.}~\bibnamefont {Behler}}, \
  and\ \bibinfo {author} {\bibfnamefont {P.}~\bibnamefont {Marquetand}},\
  }\href {https://doi.org/10.1039/C7SC02267K} {\bibfield  {journal} {\bibinfo
  {journal} {Chem. Sci.}\ }\textbf {\bibinfo {volume} {8}},\ \bibinfo {eid}
  {6924} (\bibinfo {year} {2017})}\BibitemShut {NoStop}%
\bibitem [{\citenamefont {Salazar}\ and\ \citenamefont
  {Bell}(1998)}]{Salazar_Bell:1998}%
  \BibitemOpen
  \bibfield  {author} {\bibinfo {author} {\bibfnamefont {M.~R.}\ \bibnamefont
  {Salazar}}\ and\ \bibinfo {author} {\bibfnamefont {R.~L.}\ \bibnamefont
  {Bell}},\ }\href
  {https://doi.org/10.1002/(SICI)1096-987X(199810)19:13<1431::AID-JCC1>3.0.CO;2-R}
  {\bibfield  {journal} {\bibinfo  {journal} {J. Comput. Chem.}\ }\textbf
  {\bibinfo {volume} {19}},\ \bibinfo {eid} {1431} (\bibinfo {year}
  {1998})}\BibitemShut {NoStop}%
\bibitem [{\citenamefont {Salazar}(2002)}]{Salazar:2002}%
  \BibitemOpen
  \bibfield  {author} {\bibinfo {author} {\bibfnamefont {M.~R.}\ \bibnamefont
  {Salazar}},\ }\href {https://doi.org/10.1016/S0009-2614(02)00744-3}
  {\bibfield  {journal} {\bibinfo  {journal} {Chem. Phys. Lett.}\ }\textbf
  {\bibinfo {volume} {359}},\ \bibinfo {eid} {460} (\bibinfo {year}
  {2002})}\BibitemShut {NoStop}%
\bibitem [{\citenamefont {Hornus}\ and\ \citenamefont
  {Boissonnat}(2008)}]{Hornus_Boissonnat:2008}%
  \BibitemOpen
  \bibfield  {author} {\bibinfo {author} {\bibfnamefont {S.}~\bibnamefont
  {Hornus}}\ and\ \bibinfo {author} {\bibfnamefont {J.~D.}\ \bibnamefont
  {Boissonnat}},\ }\href {https://hal.inria.fr/inria-00343188} {\enquote
  {\bibinfo {title} {An efficient implementation of {Delaunay} triangulations
  in medium dimensions},}\ } (\bibinfo {year} {2008}),\ \bibinfo {note}
  {{[Research Report] RR-6743, INRIA.}}\BibitemShut {Stop}%
\bibitem [{\citenamefont {Clough}\ and\ \citenamefont
  {Tocher}(1966)}]{Clough_Tocher:1966}%
  \BibitemOpen
  \bibfield  {author} {\bibinfo {author} {\bibfnamefont {R.~W.}\ \bibnamefont
  {Clough}}\ and\ \bibinfo {author} {\bibfnamefont {J.~L.}\ \bibnamefont
  {Tocher}},\ }in\ \href {http://contrails.iit.edu/items/show/8574} {\emph
  {\bibinfo {booktitle} {Proceedings of the Conference on Matrix Methods in
  Structural Mechanics}}}\ (\bibinfo {year} {1966})\ pp.\ \bibinfo {pages}
  {515--545}\BibitemShut {NoStop}%
\bibitem [{\citenamefont {Alfeld}(1984)}]{Alfeld:1984}%
  \BibitemOpen
  \bibfield  {author} {\bibinfo {author} {\bibfnamefont {P.}~\bibnamefont
  {Alfeld}},\ }\href {https://doi.org/10.1016/0167-8396(84)90029-3} {\bibfield
  {journal} {\bibinfo  {journal} {Comput. Aided Geom. Des.}\ }\textbf {\bibinfo
  {volume} {1}},\ \bibinfo {eid} {169} (\bibinfo {year} {1984})}\BibitemShut
  {NoStop}%
\bibitem [{\citenamefont {Alfeld}(1985)}]{Alfeld:1985}%
  \BibitemOpen
  \bibfield  {author} {\bibinfo {author} {\bibfnamefont {P.}~\bibnamefont
  {Alfeld}},\ }\href {https://doi.org/10.1137/0722006} {\bibfield  {journal}
  {\bibinfo  {journal} {SIAM J. Numer. Anal.}\ }\textbf {\bibinfo {volume}
  {22}},\ \bibinfo {eid} {95} (\bibinfo {year} {1985})}\BibitemShut {NoStop}%
\bibitem [{\citenamefont {Dell'Accio}\ \emph {et~al.}(2018)\citenamefont
  {Dell'Accio}, \citenamefont {{Di Tommaso}}, \citenamefont {Nouisser},\ and\
  \citenamefont {Zerroudi}}]{Dell'Accio_Zerroudi:2018}%
  \BibitemOpen
  \bibfield  {author} {\bibinfo {author} {\bibfnamefont {F.}~\bibnamefont
  {Dell'Accio}}, \bibinfo {author} {\bibfnamefont {F.}~\bibnamefont {{Di
  Tommaso}}}, \bibinfo {author} {\bibfnamefont {O.}~\bibnamefont {Nouisser}}, \
  and\ \bibinfo {author} {\bibfnamefont {B.}~\bibnamefont {Zerroudi}},\ }\href
  {https://doi.org/10.1016/j.apnum.2017.12.006} {\bibfield  {journal} {\bibinfo
   {journal} {Appl. Numer. Math.}\ }\textbf {\bibinfo {volume} {126}},\
  \bibinfo {eid} {78} (\bibinfo {year} {2018})}\BibitemShut {NoStop}%
\bibitem [{\citenamefont {Xuli}(2003)}]{Xuli:2003}%
  \BibitemOpen
  \bibfield  {author} {\bibinfo {author} {\bibfnamefont {H.}~\bibnamefont
  {Xuli}},\ }\href {https://doi.org/10.1016/j.jat.2003.08.001} {\bibfield
  {journal} {\bibinfo  {journal} {J. Approx. Theory}\ }\textbf {\bibinfo
  {volume} {124}},\ \bibinfo {eid} {242} (\bibinfo {year} {2003})}\BibitemShut
  {NoStop}%
\bibitem [{\citenamefont {Guessab}, \citenamefont {Nouisser},\ and\
  \citenamefont {Schmeisser}(2006)}]{Guessab_Schmeisser:2006}%
  \BibitemOpen
  \bibfield  {author} {\bibinfo {author} {\bibfnamefont {A.}~\bibnamefont
  {Guessab}}, \bibinfo {author} {\bibfnamefont {O.}~\bibnamefont {Nouisser}}, \
  and\ \bibinfo {author} {\bibfnamefont {G.}~\bibnamefont {Schmeisser}},\
  }\href {https://doi.org/10.1016/j.cam.2005.08.015} {\bibfield  {journal}
  {\bibinfo  {journal} {J. Comput. Appl. Math.}\ }\textbf {\bibinfo {volume}
  {196}},\ \bibinfo {eid} {162} (\bibinfo {year} {2006})}\BibitemShut {NoStop}%
\bibitem [{\citenamefont {Bettens}\ and\ \citenamefont
  {Collins}(1999)}]{Bettens_Collins:1999}%
  \BibitemOpen
  \bibfield  {author} {\bibinfo {author} {\bibfnamefont {R.~P.~A.}\
  \bibnamefont {Bettens}}\ and\ \bibinfo {author} {\bibfnamefont {M.~A.}\
  \bibnamefont {Collins}},\ }\href {https://doi.org/10.1063/1.479368}
  {\bibfield  {journal} {\bibinfo  {journal} {J. Chem. Phys.}\ }\textbf
  {\bibinfo {volume} {111}},\ \bibinfo {eid} {816} (\bibinfo {year}
  {1999})}\BibitemShut {NoStop}%
\bibitem [{\citenamefont {Lawson}(1986)}]{Lawson:1986}%
  \BibitemOpen
  \bibfield  {author} {\bibinfo {author} {\bibfnamefont {C.~L.}\ \bibnamefont
  {Lawson}},\ }\href {https://doi.org/10.1016/0167-8396(86)90001-4} {\bibfield
  {journal} {\bibinfo  {journal} {Comput. Aided Geom. Des.}\ }\textbf {\bibinfo
  {volume} {3}},\ \bibinfo {eid} {231} (\bibinfo {year} {1986})}\BibitemShut
  {NoStop}%
\bibitem [{\citenamefont {Edelsbrunner}(2000)}]{Edelsbrunner:2000}%
  \BibitemOpen
  \bibfield  {author} {\bibinfo {author} {\bibfnamefont {H.}~\bibnamefont
  {Edelsbrunner}},\ }\href {https://doi.org/10.1017/S0962492900001331}
  {\bibfield  {journal} {\bibinfo  {journal} {Acta Numer.}\ }\textbf {\bibinfo
  {volume} {9}},\ \bibinfo {eid} {133} (\bibinfo {year} {2000})}\BibitemShut
  {NoStop}%
\bibitem [{\citenamefont {Mirebeau}(2010)}]{Mirebeau:2010}%
  \BibitemOpen
  \bibfield  {author} {\bibinfo {author} {\bibfnamefont {J.~M.}\ \bibnamefont
  {Mirebeau}},\ }\href {https://doi.org/10.1007/s00365-010-9090-y} {\bibfield
  {journal} {\bibinfo  {journal} {Constr. Approx.}\ }\textbf {\bibinfo {volume}
  {32}},\ \bibinfo {eid} {339} (\bibinfo {year} {2010})}\BibitemShut {NoStop}%
\bibitem [{\citenamefont {Bossen}\ and\ \citenamefont
  {Heckbert}(1996)}]{Bossen_Heckbert:1996}%
  \BibitemOpen
  \bibfield  {author} {\bibinfo {author} {\bibfnamefont {F.}~\bibnamefont
  {Bossen}}\ and\ \bibinfo {author} {\bibfnamefont {P.~S.}\ \bibnamefont
  {Heckbert}},\ }in\ \href
  {https://www.semanticscholar.org/paper/A-Pliant-Method-for-Anisotropic-Mesh-Generation-Bossen-Heckbert/12318fc5882e328a330640bf38f79da5c523e1a6}
  {\emph {\bibinfo {booktitle} {5th International Meshing Roundtable}}}\
  (\bibinfo {year} {1996})\ pp.\ \bibinfo {pages} {115--126}\BibitemShut
  {NoStop}%
\bibitem [{\citenamefont {Shimada}, \citenamefont {Yamada},\ and\ \citenamefont
  {Itoh}(2000)}]{Shimada_Itoh:2000}%
  \BibitemOpen
  \bibfield  {author} {\bibinfo {author} {\bibfnamefont {K.}~\bibnamefont
  {Shimada}}, \bibinfo {author} {\bibfnamefont {A.}~\bibnamefont {Yamada}}, \
  and\ \bibinfo {author} {\bibfnamefont {T.}~\bibnamefont {Itoh}},\ }\href
  {https://doi.org/10.1142/S0218195900000243} {\bibfield  {journal} {\bibinfo
  {journal} {Int. J. Comput. Geom. Appl.}\ }\textbf {\bibinfo {volume} {10}},\
  \bibinfo {eid} {417} (\bibinfo {year} {2000})}\BibitemShut {NoStop}%
\bibitem [{\citenamefont {Boissonnat}, \citenamefont {Rouxel-Labb{\'e}},\ and\
  \citenamefont {Wintraecken}(2017)}]{Boissonnat_Wintraecken:2017}%
  \BibitemOpen
  \bibfield  {author} {\bibinfo {author} {\bibfnamefont {J.~D.}\ \bibnamefont
  {Boissonnat}}, \bibinfo {author} {\bibfnamefont {M.}~\bibnamefont
  {Rouxel-Labb{\'e}}}, \ and\ \bibinfo {author} {\bibfnamefont
  {M.}~\bibnamefont {Wintraecken}},\ }in\ \href
  {http://doi.org/10.4230/LIPIcs.SoCG.2017.19} {\emph {\bibinfo {booktitle}
  {33rd International Symposium on Computational Geometry (SoCG 2017)}}},\
  \bibinfo {series} {Leibniz International Proceedings in Informatics
  (LIPIcs)}, Vol.~\bibinfo {volume} {77},\ \bibinfo {editor} {edited by\
  \bibinfo {editor} {\bibfnamefont {B.}~\bibnamefont {Aronov}}\ and\ \bibinfo
  {editor} {\bibfnamefont {M.~J.}\ \bibnamefont {Katz}}}\ (\bibinfo
  {publisher} {Schloss Dagstuhl--Leibniz-Zentrum fuer Informatik},\ \bibinfo
  {address} {Dagstuhl, Germany},\ \bibinfo {year} {2017})\ pp.\ \bibinfo
  {pages} {19:1--19:16}\BibitemShut {NoStop}%
\bibitem [{\citenamefont {Devillers}, \citenamefont {Pion},\ and\ \citenamefont
  {Teillaud}(2002)}]{Devillers_Teillaud:2002}%
  \BibitemOpen
  \bibfield  {author} {\bibinfo {author} {\bibfnamefont {O.}~\bibnamefont
  {Devillers}}, \bibinfo {author} {\bibfnamefont {S.}~\bibnamefont {Pion}}, \
  and\ \bibinfo {author} {\bibfnamefont {M.}~\bibnamefont {Teillaud}},\ }\href
  {https://doi.org/10.1142/S0129054102001047} {\bibfield  {journal} {\bibinfo
  {journal} {Int. J. Found. Comput. S.}\ }\textbf {\bibinfo {volume} {13}},\
  \bibinfo {eid} {181} (\bibinfo {year} {2002})}\BibitemShut {NoStop}%
\bibitem [{\citenamefont {Bentley}(1975)}]{Bentley:1975}%
  \BibitemOpen
  \bibfield  {author} {\bibinfo {author} {\bibfnamefont {J.~L.}\ \bibnamefont
  {Bentley}},\ }\href {http://doi.org/10.1145/361002.361007} {\bibfield
  {journal} {\bibinfo  {journal} {Commun. ACM}\ }\textbf {\bibinfo {volume}
  {18}},\ \bibinfo {eid} {509} (\bibinfo {year} {1975})}\BibitemShut {NoStop}%
\bibitem [{\citenamefont {Flyvbjerg}\ and\ \citenamefont
  {Petersen}(1989)}]{Flyvbjerg_Petersen:1989}%
  \BibitemOpen
  \bibfield  {author} {\bibinfo {author} {\bibfnamefont {H.}~\bibnamefont
  {Flyvbjerg}}\ and\ \bibinfo {author} {\bibfnamefont {H.~G.}\ \bibnamefont
  {Petersen}},\ }\href {https://doi.org/10.1063/1.457480} {\bibfield  {journal}
  {\bibinfo  {journal} {J. Chem. Phys.}\ }\textbf {\bibinfo {volume} {91}},\
  \bibinfo {eid} {461} (\bibinfo {year} {1989})}\BibitemShut {NoStop}%
\bibitem [{\citenamefont {Herman}, \citenamefont {Bruskin},\ and\ \citenamefont
  {Berne}(1982)}]{Herman_Berne:1982}%
  \BibitemOpen
  \bibfield  {author} {\bibinfo {author} {\bibfnamefont {M.~F.}\ \bibnamefont
  {Herman}}, \bibinfo {author} {\bibfnamefont {E.~J.}\ \bibnamefont {Bruskin}},
  \ and\ \bibinfo {author} {\bibfnamefont {B.~J.}\ \bibnamefont {Berne}},\
  }\href {https://doi.org/10.1063/1.442815} {\bibfield  {journal} {\bibinfo
  {journal} {J. Chem. Phys.}\ }\textbf {\bibinfo {volume} {76}},\ \bibinfo
  {eid} {5150} (\bibinfo {year} {1982})}\BibitemShut {NoStop}%
\bibitem [{\citenamefont {Tuckerman}\ \emph {et~al.}(1993)\citenamefont
  {Tuckerman}, \citenamefont {Berne}, \citenamefont {Martyna},\ and\
  \citenamefont {Klein}}]{Tuckerman_Klein:1993}%
  \BibitemOpen
  \bibfield  {author} {\bibinfo {author} {\bibfnamefont {M.~E.}\ \bibnamefont
  {Tuckerman}}, \bibinfo {author} {\bibfnamefont {B.~J.}\ \bibnamefont
  {Berne}}, \bibinfo {author} {\bibfnamefont {G.~J.}\ \bibnamefont {Martyna}},
  \ and\ \bibinfo {author} {\bibfnamefont {M.~L.}\ \bibnamefont {Klein}},\
  }\href {https://doi.org/10.1063/1.465188} {\bibfield  {journal} {\bibinfo
  {journal} {J. Chem. Phys.}\ }\textbf {\bibinfo {volume} {99}},\ \bibinfo
  {eid} {2796} (\bibinfo {year} {1993})}\BibitemShut {NoStop}%
\bibitem [{\citenamefont {Feynman}\ and\ \citenamefont
  {Hibbs}(1965)}]{Feynman_Hibbs:1965}%
  \BibitemOpen
  \bibfield  {author} {\bibinfo {author} {\bibfnamefont {R.~P.}\ \bibnamefont
  {Feynman}}\ and\ \bibinfo {author} {\bibfnamefont {A.~R.}\ \bibnamefont
  {Hibbs}},\ }\href@noop {} {\emph {\bibinfo {title} {Quantum mechanics and
  path integrals}}}\ (\bibinfo  {publisher} {McGraw-Hill},\ \bibinfo {year}
  {1965})\BibitemShut {NoStop}%
\bibitem [{\citenamefont {P\'{e}rez}\ and\ \citenamefont {von
  Lilienfeld}(2011)}]{Perez_Lilienfeld:2011}%
  \BibitemOpen
  \bibfield  {author} {\bibinfo {author} {\bibfnamefont {A.}~\bibnamefont
  {P\'{e}rez}}\ and\ \bibinfo {author} {\bibfnamefont {O.~A.}\ \bibnamefont
  {von Lilienfeld}},\ }\href {https://doi.org/10.1021/ct2000556} {\bibfield
  {journal} {\bibinfo  {journal} {J. Chem. Theory Comput.}\ }\textbf {\bibinfo
  {volume} {7}},\ \bibinfo {eid} {2358} (\bibinfo {year} {2011})}\BibitemShut
  {NoStop}%
\bibitem [{\citenamefont {Cheng}\ and\ \citenamefont
  {Ceriotti}(2014)}]{Cheng_Ceriotti:2014}%
  \BibitemOpen
  \bibfield  {author} {\bibinfo {author} {\bibfnamefont {B.}~\bibnamefont
  {Cheng}}\ and\ \bibinfo {author} {\bibfnamefont {M.}~\bibnamefont
  {Ceriotti}},\ }\href {https://doi.org/10.1063/1.4904293} {\bibfield
  {journal} {\bibinfo  {journal} {J. Chem. Phys.}\ }\textbf {\bibinfo {volume}
  {141}},\ \bibinfo {eid} {244112} (\bibinfo {year} {2014})}\BibitemShut
  {NoStop}%
\bibitem [{\citenamefont {Makhnev}\ \emph {et~al.}(2018)\citenamefont
  {Makhnev}, \citenamefont {Kyuberis}, \citenamefont {Polyansky}, \citenamefont
  {Mizus}, \citenamefont {Tennyson},\ and\ \citenamefont
  {Zobov}}]{Makhnev_Zobov:2018}%
  \BibitemOpen
  \bibfield  {author} {\bibinfo {author} {\bibfnamefont {V.~Y.}\ \bibnamefont
  {Makhnev}}, \bibinfo {author} {\bibfnamefont {A.~A.}\ \bibnamefont
  {Kyuberis}}, \bibinfo {author} {\bibfnamefont {O.~L.}\ \bibnamefont
  {Polyansky}}, \bibinfo {author} {\bibfnamefont {I.~I.}\ \bibnamefont
  {Mizus}}, \bibinfo {author} {\bibfnamefont {J.}~\bibnamefont {Tennyson}}, \
  and\ \bibinfo {author} {\bibfnamefont {N.~F.}\ \bibnamefont {Zobov}},\ }\href
  {https://doi.org/10.1016/j.jms.2018.09.002} {\bibfield  {journal} {\bibinfo
  {journal} {J. Mol. Spectrosc.}\ }\textbf {\bibinfo {volume} {353}},\ \bibinfo
  {eid} {40} (\bibinfo {year} {2018})}\BibitemShut {NoStop}%
\bibitem [{\citenamefont {Sprik}, \citenamefont {Klein},\ and\ \citenamefont
  {Chandler}(1985{\natexlab{a}})}]{Sprik_Chandler:1985}%
  \BibitemOpen
  \bibfield  {author} {\bibinfo {author} {\bibfnamefont {M.}~\bibnamefont
  {Sprik}}, \bibinfo {author} {\bibfnamefont {M.~L.}\ \bibnamefont {Klein}}, \
  and\ \bibinfo {author} {\bibfnamefont {D.}~\bibnamefont {Chandler}},\ }\href
  {https://doi.org/10.1103/physrevb.31.4234} {\bibfield  {journal} {\bibinfo
  {journal} {Phys. Rev. B}\ }\textbf {\bibinfo {volume} {31}},\ \bibinfo {eid}
  {4234} (\bibinfo {year} {1985}{\natexlab{a}})}\BibitemShut {NoStop}%
\bibitem [{\citenamefont {Sprik}, \citenamefont {Klein},\ and\ \citenamefont
  {Chandler}(1985{\natexlab{b}})}]{Sprik_Chandler:1985_1}%
  \BibitemOpen
  \bibfield  {author} {\bibinfo {author} {\bibfnamefont {M.}~\bibnamefont
  {Sprik}}, \bibinfo {author} {\bibfnamefont {M.~L.}\ \bibnamefont {Klein}}, \
  and\ \bibinfo {author} {\bibfnamefont {D.}~\bibnamefont {Chandler}},\ }\href
  {https://doi.org/10.1103/PhysRevB.32.545} {\bibfield  {journal} {\bibinfo
  {journal} {Phys. Rev. B}\ }\textbf {\bibinfo {volume} {32}},\ \bibinfo {eid}
  {545} (\bibinfo {year} {1985}{\natexlab{b}})}\BibitemShut {NoStop}%
\bibitem [{\citenamefont {Urey}(1947)}]{Urey:1947}%
  \BibitemOpen
  \bibfield  {author} {\bibinfo {author} {\bibfnamefont {H.~C.}\ \bibnamefont
  {Urey}},\ }\href {https://doi.org/10.1039/JR9470000562} {\bibfield  {journal}
  {\bibinfo  {journal} {J. Chem. Soc.}\ }\textbf {\bibinfo {volume} {1947}},\
  \bibinfo {eid} {562} (\bibinfo {year} {1947})}\BibitemShut {NoStop}%
\bibitem [{\citenamefont {Wolfsberg}\ \emph {et~al.}(2010)\citenamefont
  {Wolfsberg}, \citenamefont {Hook}, \citenamefont {Paneth},\ and\
  \citenamefont {Rebelo}}]{Wolfsberg_Rebelo:2010}%
  \BibitemOpen
  \bibfield  {author} {\bibinfo {author} {\bibfnamefont {M.}~\bibnamefont
  {Wolfsberg}}, \bibinfo {author} {\bibfnamefont {W.~A.~V.}\ \bibnamefont
  {Hook}}, \bibinfo {author} {\bibfnamefont {P.}~\bibnamefont {Paneth}}, \ and\
  \bibinfo {author} {\bibfnamefont {L.~P.~N.}\ \bibnamefont {Rebelo}},\ }\href
  {https://doi.org/10.1007/978-90-481-2265-3} {\emph {\bibinfo {title} {Isotope
  Effects in the Chemical, Geological and Bio Sciences}}}\ (\bibinfo
  {publisher} {McGraw-Hill},\ \bibinfo {year} {2010})\BibitemShut {NoStop}%
\bibitem [{\citenamefont {Webb}\ and\ \citenamefont {{T. F. Miller
  III}}(2014)}]{Webb_Miller:2014}%
  \BibitemOpen
  \bibfield  {author} {\bibinfo {author} {\bibfnamefont {M.~A.}\ \bibnamefont
  {Webb}}\ and\ \bibinfo {author} {\bibnamefont {{T. F. Miller III}}},\ }\href
  {https://doi.org/10.1021/jp411134v} {\bibfield  {journal} {\bibinfo
  {journal} {J. Phys. Chem. A}\ }\textbf {\bibinfo {volume} {118}},\ \bibinfo
  {eid} {467} (\bibinfo {year} {2014})}\BibitemShut {NoStop}%
\bibitem [{Note1()}]{Note1}%
  \BibitemOpen
  \bibinfo {note} {Table~\ref {tab:HCN_IE_errs} shows that the largest number
  of mesh points was generated during the lowest temperature simulation.
  Because the spacing of mesh points does not depend on temperature, this
  suggests that the mesh generated at the lowest temperature covered the
  largest region of configuration space and, therefore, should be used as a
  starting mesh. This counterintuitive observation can be explained as follows:
  The lower the temperature, the greater the quantum delocalization of the ring
  polymer and the greater the explored region of configuration space. Although
  this delocalization diminishes at higher temperatures, it is eventually
  replaced by an ever increasing motion of the center of the ring polymer
  associated with the classical Boltzmann distribution. However, we cannot see
  this transition yet in Table~\ref {tab:HCN_IE_errs}, most likely because the
  quantum delocalization at lower temperatures is further increased due to the
  use of mass-scaled direct estimators, which stretch the ring polymer by a factor
  of $\protect \sqrt {2}$.\cite {Cheng_Ceriotti:2014}\protect \pseudodot
  .}\BibitemShut {Stop}%
\end{thebibliography}

\appendix

\section{Pseudo-code of the triangulation algorithm}

The pseudo-code for updating the mesh triangulation given no $D+1$ points lie in the same hyperplane is outlined in Algorithm~\ref{alg:triangulation_update}, which uses ``expand\_convex\_hull'' subroutine described in Algorithm~\ref{alg:convex_expansion}. The following notation is used:
\begin{itemize}
 \item $ \mathbf{r}_{\mathrm{add}}$ is the point being added to the mesh.
 \item We reserve $ \mathcal{S}$ for simplex variables (consisting of $ D+1$ points), $ \mathcal{F}$ for face variables (consisting of $ D$ points), $ e$ for edge variables (consisting of $D-1$ points; for $ D=2$, an edge consists of a single point, but we will still call it ``edge'' for the sake of generality), and bold font for variables that are arrays of simplices, faces, or edges.
 \item $ \bm{\mathcal{S}}_{\mathrm{mesh}}$ and $ \bm{\mathcal{F}}_{\mathrm{mesh}}$ are the current lists of simplices and faces of the convex.
\end{itemize}

\begin{algorithm}
\caption{Updating the triangulation of the mesh once a new point $ \mathbf{r}_{\mathrm{add}}$ is added.}
\label{alg:triangulation_update}
\begin{algorithmic}
 \IF{($\mathbf{r}_{\mathrm{add}}$ is inside a simplex $ \mathcal{S}\in\bm{\mathcal{S}}_{\mathrm{mesh}}$)}
 \STATE{combine each face of $ \mathcal{S}$ with $ \mathbf{r}_{\mathrm{add}}$ to create an array $ \bm{\mathcal{S}}_{\mathrm{new}}$ of $ D+1$ simplices;}
  \STATE{delete $ \mathcal{S}$;}
\ELSE
   \STATE{\textbf{call} expand\_convex\_hull($\mathbf{r}_{\mathrm{add}}$, $\bm{\mathcal{S}}_{\mathrm{new}}$);}
 \ENDIF
 \WHILE{(array $\bm{\mathcal{S}}_{\mathrm{new}}$ is not empty)}
 \STATE{choose a random $ \mathcal{S}^{\prime}\in\bm{\mathcal{S}}_{\mathrm{new}}$;}
 \STATE{delete $ \mathcal{S}^{\prime}$ from $\bm{\mathcal{S}}_{\mathrm{new}}$;}
 \FOR{\textbf{each} [$ \mathcal{S}^{\prime\prime}$ that shares a face with $ \mathcal{S}^{\prime}$ (chosen in random order)]}
 \STATE{|* Lawson flip *|}
 \STATE{form convex hull $\mathcal{C} $ from vertices of $ \mathcal{S}^{\prime\prime}$ and $ \mathcal{S}^{\prime}$;}
 \STATE{attempt to create array $ \bm{\mathcal{S}}_{\mathrm{current}}$ which triangulates $ \mathcal{C}$ with currently existing simplices;}
 \STATE{attempt to create array $ \bm{\mathcal{S}}_{\mathrm{flipped}}$ which also triangulates $ \mathcal{C}$ and differs from $ \bm{\mathcal{S}}_{\mathrm{current}}$;}
 \IF{[both $ \bm{\mathcal{S}}_{\mathrm{current}}$ and $ \bm{\mathcal{S}}_{\mathrm{flipped}}$ exist and $ G(\bm{\mathcal{S}}_{\mathrm{current}},\bm{\mathcal{S}}_{\mathrm{flipped}})>0$]}
 \STATE{delete all simplices of $\bm{\mathcal{S}}_{\mathrm{current}}$ from $ \bm{\mathcal{S}}_{\mathrm{mesh}}$ and (where present) $\bm{\mathcal{S}}_{\mathrm{new}}$;}
 \STATE{add all simplices of $ \bm{\mathcal{S}}_{\mathrm{flipped}}$ to $ \bm{\mathcal{S}}_{\mathrm{mesh}}$ and $\bm{\mathcal{S}}_{\mathrm{new}}$;}
 \STATE{exit the \textbf{for} loop;}
 \ENDIF
 \ENDFOR
 \ENDWHILE
\end{algorithmic}
 \end{algorithm}

 \begin{algorithm}
\caption{Procedure for expanding $ \mathcal{C}_{\mathrm{mesh}}$ once a new point $ \mathbf{r}_{\mathrm{add}}$ is added outside of it.}
\label{alg:convex_expansion}
\begin{algorithmic}
\PROCEDURE{expand\_convex\_hull($\mathbf{r}_{\mathrm{add}}$, $ \bm{\mathcal{S}}_{\mathrm{new}}$)}
\STATE{find a face $ \mathcal{F}_{\mathrm{start}}$ such that $ n_{\mathcal{F}_{\mathrm{start}}}(\mathbf{r}_{\mathrm{add}})<0$;}
\STATE{|* $ \bm{\mathcal{F}}_{\mathrm{conf}}$ will consist of all ``conflicting'' faces $ \mathcal{F}$ such that $ n_{\mathcal{F}}(\mathbf{r}_{\mathrm{add}})<0$, $\mathbf{e}_{\mathrm{bound}} $ will consist of edges of $\bm{\mathcal{F}}_{\mathrm{conf}}$  not shared by two faces in the array *|}
\STATE{\textbf{call} expand\_conflict\_zone($\mathbf{r}_{\mathrm{add}} $, $\mathcal{F}_{\mathrm{start}}$, $\bm{\mathcal{F}}_{\mathrm{conf}}$, $ \mathbf{e}_{\mathrm{bound}}$)}
\FOR{\textbf{each} ($\mathcal{F}\in\bm{\mathcal{F}}_{\mathrm{conf}}$)}
  \STATE{create a simplex $ \mathcal{S}$ from $ \mathbf{r}_{\mathrm{add}}$ and $\mathcal{F}$;}
  \STATE{add $ \mathcal{S}$ to the current triangulation and $ \bm{\mathcal{S}}_{\mathrm{new}}$;}
  \ENDFOR
\FOR{\textbf{each} ($e\in \bm{e}_{\mathrm{bound}}$)}
  \STATE{create a face $ \mathcal{F}$ from $ \mathbf{r}_{\mathrm{add}}$ and $e$;}
  \STATE{add $ \mathcal{F}$ to $ \bm{\mathcal{F}}_{\mathrm{mesh}}$;}
  \ENDFOR
\ENDPROCEDURE
 \end{algorithmic}
\begin{algorithmic}
\PROCEDURE[\textbf{recursive }]{expand\_conflict\_zone($\mathbf{r}_{\mathrm{add}}$, $ \mathcal{F}_{\mathrm{in}}$, $ \bm{\mathcal{F}}_{\mathrm{conf}}$, $ \bm{e}_{\mathrm{bound}}$)}
\STATE{add $\mathcal{F}_{\mathrm{in}}$ to $ \bm{\mathcal{F}}_{\mathrm{conf}}$;}
\FOR{\textbf{each} ($\mathcal{F}^{\prime}$ sharing an edge $ e$ with $ \mathcal{F}_{\mathrm{in}}$)}
  \STATE{\textbf{if} ($\mathcal{F}^{\prime}\in\bm{\mathcal{F}}_{\mathrm{conf}}$) cycle}
  \IF{[$ n_{\mathcal{F}^{\prime}}(\mathbf{r}_{\mathrm{add}})<0$]}
  \STATE{\textbf{call} expand\_conflict\_zone($\mathbf{r}_{\mathrm{add}}$, $ \mathcal{F}^{\prime}$, $ \bm{\mathcal{F}}_{\mathrm{conf}}$)}
  \ELSE
  \STATE{add $e$ to $ \bm{e}_{\mathrm{bound}}$;}
  \ENDIF
  \ENDFOR
\ENDPROCEDURE
\end{algorithmic}

 \end{algorithm}

Algorithm~\ref{alg:convex_expansion} is quite similar to the ``conflict zone'' procedure for updating a Delaunay triangulation once a new point is added.\cite{Hornus_Boissonnat:2008} The latter  completes all convex faces with a virtual point at infinity, thus creating ``virtual simplices'' and constructs a ``conflict zone'' from simplices that have a ``conflict'' with the new point by containing it inside their circumpshere (for virtual simplices this means that the point is in the half-space on the other side of the face's hyperplane than the rest of the mesh points); the conflict zone is then replaced with new simplices. By constraining this algorithm to include in the conflict zone only virtual simplices one obtains the ``expand\_convex\_hull'' subroutine. The concept of ``virtual simplices'' from Ref.~\onlinecite{Hornus_Boissonnat:2008} can be used to avoid using a separate routine for expanding the convex hull. In this case the virtual simplex corresponding to $ \mathcal{F}_{\mathrm{start}}$ can be considered the one containing the new point $ \mathbf{r}_{\mathrm{add}}$, and the algorithm can proceed directly to Lawson flips with the definition of $ G$ [see Eq.~(\ref{eq:flip_functional})] extended to cases when virtual simplices are included into the sum. We still use the ``expand\_convex\_hull'' subroutine to make the triangulation computationally cheaper, but we will use the notion of virtual simplices a little later.

As mentioned in Sec.~\ref{sec:Theory}, it is beneficial to place some mesh points on constraining surfaces~(\ref{eq:constraining_surface}). In this case each point of the mesh is assigned a logical variable array whose $ i$th element indicates whether the point lies in the constraining plane with index $ i$. To account for constraints, the following modifications should be made to Algorithms~\ref{alg:triangulation_update}-\ref{alg:convex_expansion}:
\begin{enumerate}
 \item Each evaluation of $ n_{\mathcal{F}}(\mathbf{r}_{\mathrm{add}})$ should be preceded by checking whether a face $\mathcal{F}$ and $ \mathbf{r}_{\mathrm{add}}$ lie in the same constraining plane; if this is the case $ n_{\mathcal{F}}(\mathbf{r}_{\mathrm{add}})$ is considered exactly zero. If the face $\mathcal{F}$ does not lie in a single constraining plane in the first place, the logical check should return false.
 \item Each time $ G(\bm{\mathcal{S}}_{\mathrm{current}})$ and $G(\bm{\mathcal{S}}_{\mathrm{flipped}})$ are evaluated, it should be checked whether $ \bm{\mathcal{S}}_{\mathrm{flipped}}$ contains any simplex whose volume is zero (for example, if all of the vertices lie in a single hyperplane). If one of the faces of this simplex has zero area, the flip is considered impossible. Otherwise, faces of such simplices are turned into virtual simplices to be added to $ \bm{\mathcal{S}}_{\mathrm{current}}$ if they are already present in the triangulation, and to $ \bm{\mathcal{S}}_{\mathrm{flipped}}$ otherwise. Such virtual simplices are not added to $ \bm{\mathcal{S}}_{\mathrm{new}}$ after a successful Lawson flip and we consider their volume to be zero while evaluating the finalized $ G(\bm{\mathcal{S}}_{\mathrm{current}})$ and $G(\bm{\mathcal{S}}_{\mathrm{flipped}})$.
\end{enumerate}

\end{document}